\documentclass[trackchanges,twocolumn]{aastex701}

\usepackage{graphicx,subcaption}
\usepackage{amsmath}
\UseRawInputEncoding
\usepackage{fix-cm}
\usepackage[T1]{fontenc}
\usepackage{newtxtext,newtxmath}
\usepackage{balance}
\usepackage{xcolor}
\usepackage{placeins}

\begin{document}

\title{Probing the High-Energy Emission of the VHE-emitting Changing-Look Blazar B2~1420+32}

\author[orcid=0009-0003-7923-7624]{Anjum Peer}
\affiliation{Department of Physics, Islamic University of Science and Technology, Kashmir 192122, India}
\email[show]{peer.anjum@iust.ac.in}

\author[orcid=0000-0003-1458-4396]{Zahir Shah}
\affiliation{Manipal Centre for Natural Sciences, Centre of
Excellence, Manipal Academy of Higher Education, Manipal 576104, India}
\email[show]{zahir.shah@manipal.edu}

\author[orcid=0009-0005-3498-1104]{Sikandar Akbar}
\affiliation{Department of Physics, University of Kashmir, Srinagar 190006, India}
\email{darprince46@gmail.com}

\author[orcid=0000-0002-0360-1851]{Bari Maqbool}
\affiliation{Department of Physics, Islamic University of Science and Technology, Kashmir 192122, India}
\affiliation{Inter-University Centre for Astronomy and Astrophysics, Post Bag 4, Ganeshkhind, Pune 411007, India}
\email[show]{bmaqbool@iust.ac.in}

\author[orcid=0000-0002-7609-2779]{Ranjeev Misra}
\affiliation{Inter-University Centre for Astronomy and Astrophysics, Post Bag 4, Ganeshkhind, Pune 411007, India}
\email{ranjeev@iucaa.in}


\begin{abstract}
We present a multi-wavelength temporal and spectral study of the changing-look blazar B2~1420+32 using \emph{Fermi}-LAT, \emph{Swift}-XRT, and \emph{Swift}-UVOT data from MJD~58818--60721. The source reached a peak 0.1--300~GeV photon flux of $(4.62 \pm 0.29) \times 10^{-6}\,\mathrm{ph\,cm^{-2}\,s^{-1}}$ around MJD~60488, about 60 times the 4FGL-DR4 average, during which the photon index hardened to $2.19 \pm 0.14$; the flux--index evolution shows only weak evidence for global harder-when-brighter behaviour. The fractional variability is strongly energy dependent, largest in $\gamma$-rays, substantial in the optical/UV, and low in X-rays. Strong $\gamma$-ray--optical/UV correlations and a moderate $\gamma$-ray--X-ray correlation are observed. 
We identified five activity states for broadband spectral energy distribution (SED) modelling. Their X-ray spectra are described by a log-parabola, with significant
negative curvature in four of the five states. The high-energy (X-ray and $\gamma$-ray) components of  their SED is modeled using synchrotron self-Compton (SSC), external Compton (EC), and SSC+EC scenarios. The SSC-only and EC-only models either require physically disfavoured parameters or fail to reproduce the VHE emission, whereas SSC+EC provides the most self-consistent description, with a seed-photon temperature of $\sim 10^{3}$~K favouring an infrared torus origin. The brighter states require larger bulk Lorentz factors and higher jet powers, while the magnetic field is comparable across most states but drops sharply in the VHE state.  The break energy also changes significantly between outbursts, indicating that the spectral shape evolves between activity episodes. These results suggest that Doppler
boosting and jet energetics set the activity level, the break energy
governs the spectral shape between outbursts.
\end{abstract}

\section{Introduction}

Active galactic nuclei (AGN) are among the most luminous persistent sources in the Universe, powered by accretion of matter onto supermassive black holes. A fraction of AGN are radio-loud and launch powerful, collimated relativistic jets. Their observed properties depend strongly on the angle between the jet axis and the observer's line of sight. When the jet is closely aligned with the line of sight, relativistic beaming amplifies the jet emission, and the source is observed as a blazar \citep{1995PASP..107..803U}. Blazars are dominated by non-thermal radiation across the electromagnetic spectrum and are characterized by rapid flux and spectral variability. Their broadband spectral energy distribution (SED), typically represented in the $\nu$--$\nu F_{\nu}$ plane, exhibits a characteristic double-peaked structure \citep{2010ApJ...716...30A}. The low-energy component, extending from radio to X-ray energies, is generally attributed to synchrotron radiation from relativistic electrons within the jet \citep{1992ApJ...397L...5M, 1993ApJ...407...65G}. The origin of the high-energy component, which peaks in the MeV--GeV $\gamma$-ray regime and can extend to TeV energies in some sources, is still debated. In leptonic scenarios, the high-energy emission is produced through inverse Compton (IC) scattering, in which relativistic electrons upscatter low-energy photons to $\gamma$-ray energies. The seed photons may be synchrotron photons produced within the jet, giving rise to SSC emission \citep{1992ApJ...397L...5M, 1989ApJ...340..181G}, or photons from external radiation fields, giving rise to external Compton (EC) emission. These external photons may originate from the accretion disc, broad-line region (BLR), or dusty torus \citep{1993ApJ...416..458D, boettcher1997gammarayemissionspectralevolution}. While leptonic models successfully reproduce the observed SEDs of many blazars, alternative explanations involving hadronic or lepto-hadronic processes have also been proposed. In these scenarios, the high-energy emission may arise from proton synchrotron radiation or from electromagnetic cascades initiated by proton--photon or proton--proton interactions \citep{AHARONIAN2000377, mannheim1993protonblazar}.

Blazars are broadly classified into two subclasses: flat-spectrum radio quasars (FSRQs) and BL~Lacertae (BL~Lac) objects. Traditionally, this classification is based on the rest-frame equivalent width (EW) of the strongest broad emission line, with BL~Lacs defined by EW $< 5$\,\AA\ and FSRQs by EW $> 5$\,\AA\ \citep{1995PASP..107..803U}. However, this observational criterion can be affected by variability of the beamed jet continuum, which may dilute the observed emission lines. A more physical classification has therefore been proposed using the broad-line region luminosity normalized to the Eddington luminosity, with a dividing value of \(L_{\rm BLR}/L_{\rm Edd} \sim 5\times10^{-4}\) \citep{10.1111/j.1365-2966.2011.18578.x}. Blazars are also classified according to the synchrotron peak frequency as low-, intermediate-, and high-synchrotron-peaked sources,  hereafter referred to as LSPs, ISPs, and HSPs, respectively, with LSPs having \(\nu_{\rm peak} < 10^{14}\) Hz, ISPs having \(10^{14} < \nu_{\rm peak} < 10^{15}\) Hz, and HSPs having \(\nu_{\rm peak} > 10^{15}\) Hz \citep{2010ApJ...716...30A}. FSRQs are generally associated with the LSP class. 

FSRQs are prominent emitters in the GeV $\gamma$-ray band; however, they have been detected far less frequently in the very-high-energy (VHE; $E > 100$\,GeV) regime than BL~Lac objects \citep{TeVCat}. This scarcity is generally attributed to several factors. First, the high-energy component in FSRQs typically peaks at lower energies, resulting in a steeply falling spectrum in the VHE regime. In addition, VHE photons produced within the central regions can undergo strong internal absorption through interactions with the dense photon field of the broad-line region (BLR; \citealt{2006ApJ...653.1089L}). Their detectability is further reduced by attenuation due to the extragalactic background light (EBL), particularly because FSRQs are often located at relatively high redshifts \citep{2011MNRAS.410.2556D}. Consequently, most VHE detections of FSRQs have been obtained during flaring states.

Although blazars are traditionally classified as either FSRQs or BL~Lac objects, some sources appear to move between these observational states. More broadly, changing-look AGN are systems that undergo pronounced changes in their observed spectral properties \citep[e.g.][]{Matt_2003,Mishra_2021}. For blazars, this phenomenon is of particular interest because variability in the beamed jet emission can strongly alter the prominence of the broad emission lines. Consequently, transitions between FSRQ-like and BL~Lac-like states may reflect changes in the relative dominance of the jet and accretion-flow components, rather than orientation alone. Such sources offer a useful probe of the coupling between accretion processes, line-emitting regions, and relativistic jet activity \citep{1995A&A...293..665F}.

The flat-spectrum radio quasar B2~1420+32 (OQ~334; QSO~B1420+326) is a high-redshift blazar at \(z = 0.682\) \citep{2010MNRAS.405.2302H,2007ApJS..171...61H}. It has emerged as an important source for investigating the connection between jet emission, BLR properties, and blazar state transitions. After remaining in a relatively low state during much of the early \textit{Fermi}-LAT era, the source entered a phase of pronounced activity from 2018 onward, showing strong optical and \(\gamma\)-ray brightening together with repeated transitions between FSRQ-like and BL~Lac-like states \citep{2021ApJ...913..146M}. Spectroscopic monitoring suggests that these classification changes are driven primarily by large variations in the jet continuum, which dilute the broad emission lines \citep{2021ApJ...913..146M}. The source attracted further attention when the MAGIC telescopes detected VHE \(\gamma\)-ray emission during a high state in January 2020, making B2~1420+32 one of the most distant blazars detected at VHE energies and an important probe of particle acceleration and radiative processes in powerful jets at large cosmological distances \citep{ATel13412}.
Since the 2020 event, B2~1420+32 has undergone several episodes of enhanced activity across the electromagnetic spectrum, reinforcing its status as a highly variable FSRQ.


A prominent outburst was observed in late June 2024, when the \textit{Fermi}-LAT
reported renewed GeV \(\gamma\)-ray activity from the source
\citep{2024ATel16680....1C}. The preliminary analysis reported a daily averaged
photon flux (\(E > 100\,\mathrm{MeV}\)) of \((4.4 \pm 0.3)\times10^{-6}\)
photons\,cm\(^{-2}\)\,s\(^{-1}\) (statistical uncertainty only) on 2024 June 27
(MJD 60488), corresponding to an enhancement of about a factor of 60 relative to
the average flux listed in the 4FGL-DR4 catalogue. This exceeded the highest daily
flux of the previous \textit{Fermi}-LAT flaring episode, when the source reached
\((1.7 \pm 0.2)\times10^{-6}\) photons\,cm\(^{-2}\)\,s\(^{-1}\) on 2019
December 31 \citep{2020ATel13382....1C}, by a factor of about 2.5. The adjacent
days were substantially fainter, with \((0.9 \pm 0.1)\times10^{-6}\) and
\((1.8 \pm 0.2)\times10^{-6}\) photons\,cm\(^{-2}\)\,s\(^{-1}\) on June 26 and
28 respectively, indicating that the outburst developed and decayed on
timescales of about a day. Structure on shorter timescales is also evident: the
peak 6-hour integrated flux (\(E > 100\,\mathrm{MeV}\)) reached
\((6.5 \pm 0.7)\times10^{-6}\) photons\,cm\(^{-2}\)\,s\(^{-1}\) in the
06:00--12:00 UT interval on June 27 \citep{2024ATel16680....1C}, exceeding the
daily average by roughly 50 per cent and pointing to a compact emission region
and efficient particle acceleration within the jet. At the time of that report
this was the fourth episode of enhanced \(\gamma\)-ray activity announced from
this blazar by the \textit{Fermi}-LAT Collaboration \citep{2018ATel12277....1C,
2019ATel12942....1A, 2020ATel13382....1C, 2024ATel16680....1C}. Our independent
analysis of this flare (Section~\ref{temporal}) yields a peak one-day flux
consistent with this reported value.

Contemporaneous observations revealed a clear multiwavelength counterpart to the
2024 \(\gamma\)-ray flare. Optical monitoring showed pronounced brightening:
against a typical inter-flare level of \(R \sim 19\) mag, the source rose to
\(R = 16.53\) by 2024 May 13 and reached \(R = 14.35\)--14.78 between 2024 June
25 and 29 (MJD 60486--60490), bracketing the \(\gamma\)-ray peak
\citep{Brown2024ATel16681}; the brightest optical measurement was recorded on
June 29, two days after the \(\gamma\)-ray maximum. Multi-band photometry from
the Indian Astronomical Observatory showed the source still in an elevated state
through late July \citep{atel16782}. The optical brightening
nevertheless remained fainter than the \(R = 13.70\) mag reached during the 2019
outburst \citep{Marchini2019ATel12914}, even though the 2024 event produced the
highest daily \(\gamma\)-ray flux reported from this source up to that time.
Together with the flaring episodes of 2019 and 2020
\citep{2019ATel12886....1M, Minev2020ATel13421, DAmmando2020ATel13428}, these
observations demonstrate the recurrent nature of extreme activity in
B2~1420+32.

These properties motivate a detailed multiwavelength investigation across
different activity states, in order to understand the origin of the VHE emission
and the physical drivers of the changing-look behaviour of this source. In this
work, we present a multiwavelength temporal and spectral study of B2~1420+32
covering MJD 58818--60721, an interval that includes both the VHE-detected state
of 2020 January and the 2024 outburst.
The changing-look behaviour of B2~1420+32 has been established spectroscopically, through the decrease in the equivalent widths of its broad emission lines as the jet continuum brightens \citep{2021ApJ...913..146M,Mishra_2021}. This classification was independently supported during the 2020 outburst, when the Mg\,II equivalent width fell below the conventional threshold separating the two classes at the brightest epoch \citep{2021A&A...647A.163M}. The present study does not include emission-line spectroscopy and therefore does not independently assess or track the changing-look classification. Instead, we characterize the state-dependent broadband continuum and show that its spectral shape is not preserved between outbursts, with the relative contributions of the internal and external radiation fields changing with source state. While the subsequent activity of the source has not been examined in detail, our results provide a broadband view of the spectral evolution accompanying its changing activity. We identify five
states with simultaneous \(\gamma\)-ray, X-ray, and optical/UV coverage,
spanning 2019 December to 2024 August, and model each within a one-zone leptonic
framework implemented as a local convolution model in \textsc{xspec}
\citep{ 2024MNRAS.527.5140S,2024ApJ...977..111A}. Rather
than adopting a combined scenario at the outset, we test pure SSC, pure EC, and
combined SSC+EC descriptions against each state in turn, allowing a direct
assessment of which radiative process accounts for the observed VHE emission and
of how the inferred physical conditions evolve between outbursts separated by
several years.

The paper is organised as follows: in Section~\ref{observation}, we describe the
observations and data reduction procedures; in Section~\ref{temporal}, we
analyse the temporal variability; in Section~\ref{broadband}, we perform
detailed broadband SED modelling; in Section~\ref{summary}, we present our main
results; and in Section~\ref{discussion}, we discuss their implications for the
emission mechanisms and the changing-look nature of the source. We adopt a flat $\Lambda$CDM cosmology with $\Omega_{\Lambda} = 0.73$, $\Omega_{m} = 0.27$, and $H_{0} = 71~\mathrm{km\,s^{-1}\,Mpc^{-1}}$ throughout this work.

\section{Observations and Data Analysis}\label{observation}
We performed a detailed multi-wavelength temporal and spectral analysis of B2~1420+32 using observations from \textit{Fermi}-LAT, \textit{Swift}-XRT, and \textit{Swift}-UVOT covering the period MJD~58818--60721 (2019 December 01 to 2025 February 15). B2~1420+32 has attracted particular attention in recent years because of its pronounced high-energy activity and its detection at VHE \(\gamma\)-rays by the MAGIC telescopes in January 2020. The source was detected with a significance exceeding \(13\sigma\) during a 1.6~hr observation, with a flux corresponding to about 15\% of the Crab Nebula flux above 100~GeV \citep{Mirzoyan2020}. This VHE detection was triggered by enhanced GeV emission reported by \textit{Fermi}-LAT and was followed by multi-wavelength observations. Motivated by the pronounced variability exhibited by the source across the electromagnetic spectrum, we investigate its temporal and spectral behaviour using \textit{Fermi} and \textit{Swift} observations spanning different activity states. The details of the observations and data reduction procedures are presented below.

\subsection{Fermi-LAT}
\noindent
The \textit{Fermi} Large Area Telescope (LAT) is a pair-conversion $\gamma$-ray detector operating in the energy range from 20 MeV to above 300 GeV \citep{Atwood_2009}. 
In its standard sky-survey mode, the LAT scans the entire sky approximately every three hours.
In this work, we analyze LAT observations of B2~1420+32 obtained during the period MJD 58818-60721 in the energy range 0.1--300 GeV. The data were processed using the \texttt{Fermitools} software (version 2.2.0) following the standard analysis procedures recommended by the \textit{Fermi}-LAT collaboration. We selected P8R3 SOURCE-class events within a 15$^\circ$ region of interest (ROI) centered on the source position, using the event selections \texttt{evclass = 128} and \texttt{evtype = 3} to retain high-confidence $\gamma$-ray photons. To minimize contamination from Earth limb emission, events with zenith angles greater than 90$^\circ$ were excluded. 
The source model was constructed using the 4FGL catalog \citep{Abdollahi_2020}, including all cataloged sources within 15$^\circ$ of the target position together with their associated spectral models and catalog parameter values. Sources within 10$^\circ$ of B2~1420+32 were allowed to vary in normalization and, where appropriate, in the relevant spectral parameters, while the parameters of sources beyond 10$^\circ$ were fixed to their catalog values. The Galactic diffuse emission was modeled using \texttt{gll\_iem\_v07.fits}, while the isotropic background component was described by \texttt{iso\_P8R3\_CLEAN\_V3\_v1.txt}. The appropriate instrument response function (IRF), \texttt{P8R3\_SOURCE\_V3}, was applied.
Finally, for SED modeling, the LAT spectral points were converted into XSPEC-readable PHA format using the tool \texttt{ftflx2xsp}.

\subsection{Swift-XRT}

X-ray observations of B2~1420+32 were carried out using the Swift-XRT instrument on board the Neil Gehrels Swift Observatory \citep{2004ApJ...611.1005G}. During the period MJD~58818--60721, a total of 54 observations were obtained
(see Table~\ref{tab:b2_obs}). The photon-counting mode data were processed
using the XRTDAS v3.7.0 software package within the HEASOFT framework
(version 6.32.1) \citep{HEASoft2014}. Cleaned level 2 event files were produced following standard procedures using \texttt{XRTPIPELINE} (version 0.13.7). For spectral extraction, source events were selected from a circular region of 50 arcseconds radius, while background events were taken from a surrounding circular source free  region of 100 arcseconds radius. Exposure maps were generated with \texttt{XIMAGE}, and auxiliary response files were created using the \texttt{xrtmkarf} task. The source spectra were grouped using \texttt{GRPPHA} so that each bin contained at least 20 counts. Spectral modeling was carried out in \texttt{XSPEC} version 12.13.1, fitting either a power-law or log-parabola model and including absorption due to neutral hydrogen using the \texttt{Tbabs} model. The neutral hydrogen column density (nH) was fixed at $1.07\times10^{20}~{\rm cm^{-2}}$ \citep{HI4PI2016}, while normalization and spectral parameters were allowed to vary freely to achieve the best-fit results.

\subsection{Swift-UVOT}\label{uvot}

In addition to X-ray observations, the \textit{Swift} observatory also provides optical and ultraviolet data through the UVOT instrument \citep{Roming_2005}. The UVOT covers both optical and UV energy bands using six filters: three in the optical (U, B, V) and three in the UV (UVW1, UVM2, UVW2) \citep{10.1111/j.1365-2966.2007.12563.x}. For B2~1420+32, the UVOT data were processed with the HEASOFT software package (version~6.32.1). Image processing was carried out using the \texttt{UVOTSOURCE} task, while Multiple exposures in individual filters were first combined using \texttt{UVOTIMSUM}, and source photometry was then extracted with \texttt{UVOTSOURCE}. Source counts were extracted from a circular region of 5 arcsecond radius centered on the target, 
with a nearby circular background region of 10  arcsecond radius. To account for Galactic extinction, corrections were applied following 
\citet{Schlafly_2011}, adopting values of \(E\left(B-V\right) = 0.010\) and $R_{V} = A_{V}/E\left(B-V\right) = 3.1$. The resulting UVOT flux measurements and corresponding energies were finally converted into PHA files using the \texttt{ftflx2xsp} tool.

\section{Temporal Analysis} 
\label{temporal}

\begin{figure*}
    \centering
    \includegraphics[width=1.1\textwidth]{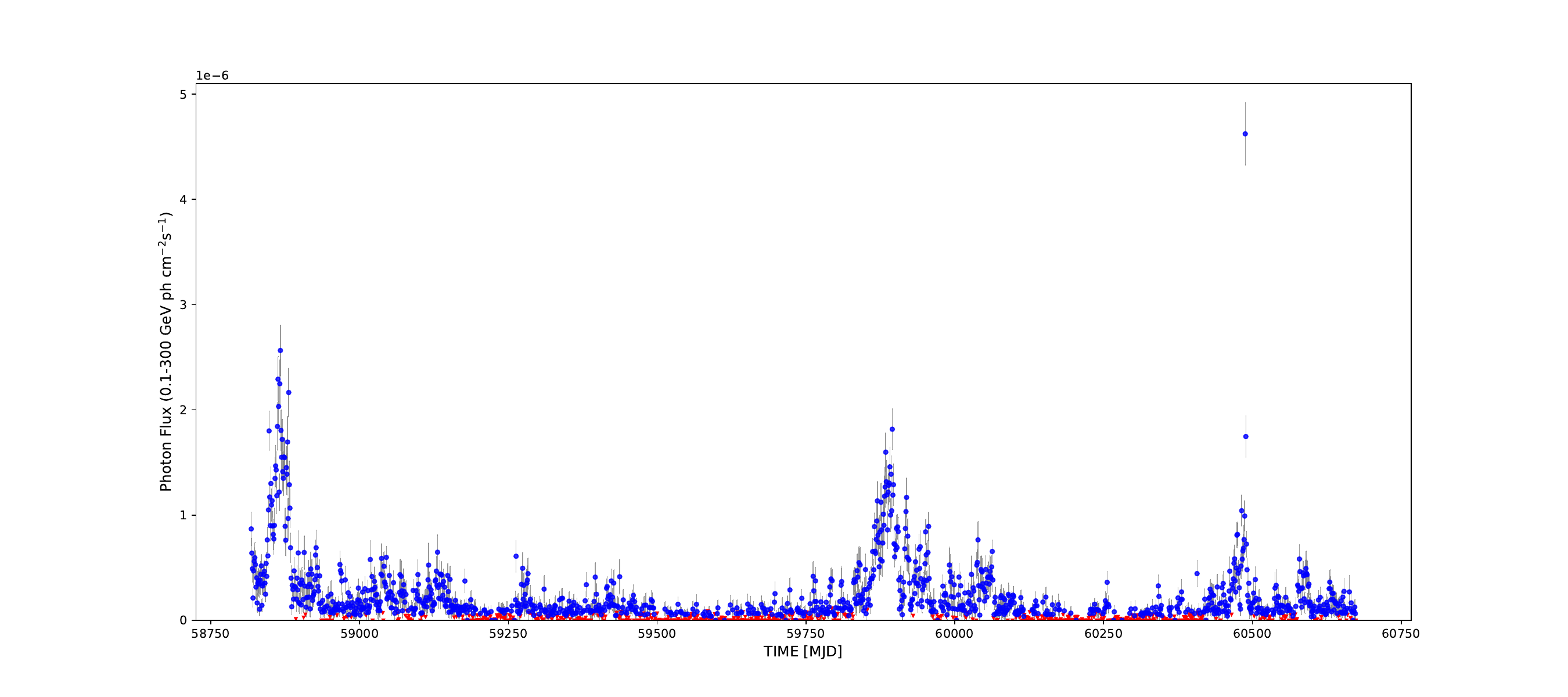}
    \caption{One-day binned $\gamma$-ray light curve of B2 1420+32 in the energy range 0.1--300\,GeV, between MJD 58818--60721, red inverted triangles indicate data points where the flux uncertainty is greater than the measured flux}
    \label{fig:fermi_lc}
\end{figure*}

Owing to the pronounced activity of B2~1420+32 observed across multiple energy bands during the interval MJD~58818--60721, we commenced temporal analysis with data from the \textit{Fermi}-LAT. The all sky monitoring capability of \textit{Fermi}-LAT makes it  suitable for studying the long-term variability of this source. To investigate the temporal characteristics, we produced a one-day binned $\gamma$-ray light curve spanning the period MJD~58818--60721, in Figure~\ref{fig:fermi_lc}.

The one-day binned \(\gamma\)-ray light curve shows a prominent flare with a
peak flux of \((4.62 \pm 0.29)\times10^{-6}\) ph\,cm\(^{-2}\)\,s\(^{-1}\) on
MJD 60488 (2024 June 27), detected with \(TS = 1754\). This is consistent within uncertainties with the
preliminary value of \((4.4 \pm 0.3)\times10^{-6}\) ph\,cm\(^{-2}\)\,s\(^{-1}\)
reported for the same day by \citet{2024ATel16680....1C}. During this high-activity state, the photon index reached a hard value of $(2.19 \pm 0.14)$, suggesting enhanced high-energy emission during the flare. To investigate the overall spectral evolution, we examined the correlation between the \(\gamma\)-ray flux and photon index across the one-day binned light curve during the interval MJD~58818--60721. A Spearman rank analysis yields \(\rho = -0.20\) with a null-hypothesis probability of \(9.5\times10^{-5}\). The anti-correlation is therefore statistically significant, but weak in amplitude, indicating that any global harder-when-brighter behaviour in B2~1420+32 is mild.

\begin{figure*}
    \centering
    \includegraphics[width=\textwidth]{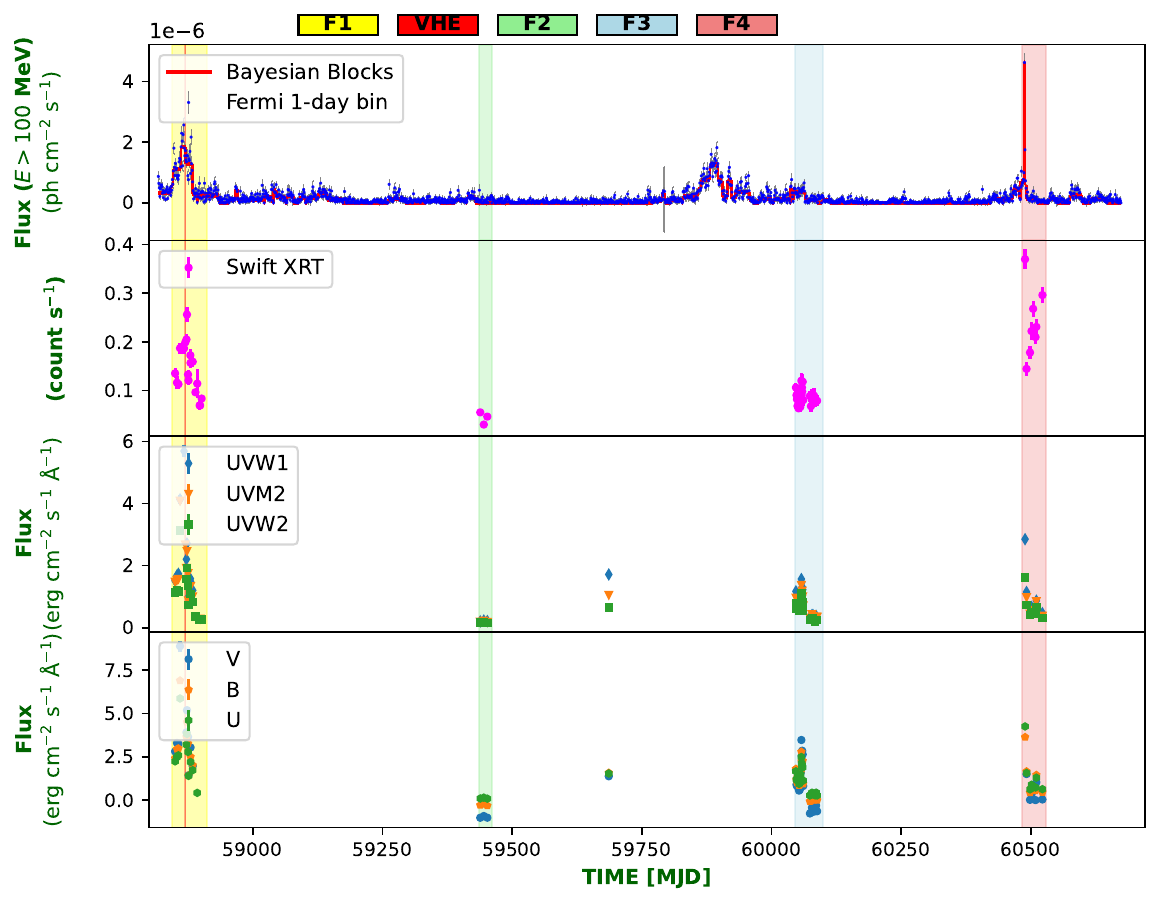}
    \caption{Multi-wavelength light curve of B2 1420+32 in different flux states.
    The top panel displays the 1-day binned $\gamma$-ray light curve integrated over 0.1--300\,GeV, including data points with $\mathrm{TS}>4$.
    The upper middle panel displays the X-ray light curve in 0.3--10\,keV.
    The lower and bottom panels display the optical and UV light curves, respectively.}
    \label{fig:mwl}
\end{figure*}

To characterize the multi-wavelength variability of B2 1420+32 during MJD~58818--60721,  we constructed multiwavelength light curves as illustrated in Figure~\ref{fig:mwl}. This specific interval was selected on the basis of alerts from the ATel \citep{Anand2024ATel16782,  ATel17504}, which highlighted episodes of elevated activity across several energy bands, particularly in the UVOT and high energy $\gamma$-ray domains. In addition, the availability of contemporaneous observations from multiple instruments during this period further enabled a detailed investigation of the source behaviour across the electromagnetic spectrum.
The top panel of Figure~\ref{fig:mwl} presents the $\gamma$-ray light curve in the 0.1--300~GeV range, derived from \textit{Fermi}-LAT data and binned over one day. The subsequent panels display results from \textit{Swift} observations: the upper middle panel shows the X-ray light curve from the XRT, the lower middle panel depicts the optical bands measured with UVOT, and the bottom panel illustrates the UV bands from the same instrument. Each point in the X-ray and UV/optical light curves corresponds to an individual observation ID retrieved from the \textit{Swift} archive.
These multiwavelength light curves provide a coherent picture of the variability exhibited by B2~1420+32 and enable a direct comparison of flux changes across different regions of the electromagnetic spectrum.
The light curves reveal pronounced variability across the optical/UV, X-ray and
\(\gamma\)-ray bands. During MJD 58818--60721, the one-day binned
\(\gamma\)-ray light curve reaches a peak integrated flux of
\((4.62 \pm 0.29)\times10^{-6}\) ph\,cm\(^{-2}\)\,s\(^{-1}\) at MJD 60488, and
the X-ray flux is also strongly enhanced around this epoch, reaching
\(0.37 \pm 0.02\) counts\,s\(^{-1}\). The optical/UV emission, however, attains
its highest level of \((8.82 \pm 0.31)\times10^{-14}\)
erg\,cm\(^{-2}\)\,s\(^{-1}\)\,\AA\(^{-1}\) in the V band at MJD 58859,
during the 2020 outburst, rather than during the brightest \(\gamma\)-ray flare
of 2024. The two outbursts therefore differ in their relative band amplitudes: the 2020
event reached the higher optical/UV level, while the 2024 event produced the
higher \(\gamma\)-ray flux. This suggests that the broadband spectral shape is
not preserved from one outburst to the next, and that the relative output of the
synchrotron and inverse-Compton components differs between episodes of enhanced
activity. A related result was obtained for the 2020 outburst by
\citet{2021A&A...647A.163M}, who found that although the source remained Compton
dominated, as is typical for FSRQs, the degree of dominance at the peak of that
flare was only a factor of a few. We note that the present comparison
concerns the peak amplitudes of two outbursts separated by more than four years
and does not constitute a simultaneous measurement of the two components; it is
distinct from the epoch-by-epoch correlation analysis presented below, which
quantifies how closely the bands track one another within the sampled epochs.

To further assess the variability characteristics evident in the multiwavelength light curves, we estimated the fractional variability amplitude ($F_{\mathrm{var}}$) across the different energy bands.  
For this purpose, only flux measurements that were simultaneously observed among the various bands were considered, ensuring a consistent comparison of variability on the same temporal scales. This approach enables examination of how the strength of variability changes with photon energy. The computation of $F_{\mathrm{var}}$ was carried out following the formalism proposed by \citet{2003MNRAS.345.1271V}, where the variability amplitude is determined using the expression:
 \begin{equation}
\rm F_{var}=\sqrt{\frac{S^2-\overline{\sigma^{2}_{err}}}{\overline{F}^2}}
\end{equation}
\begin{figure}
	\centering
	\includegraphics[scale=0.35]{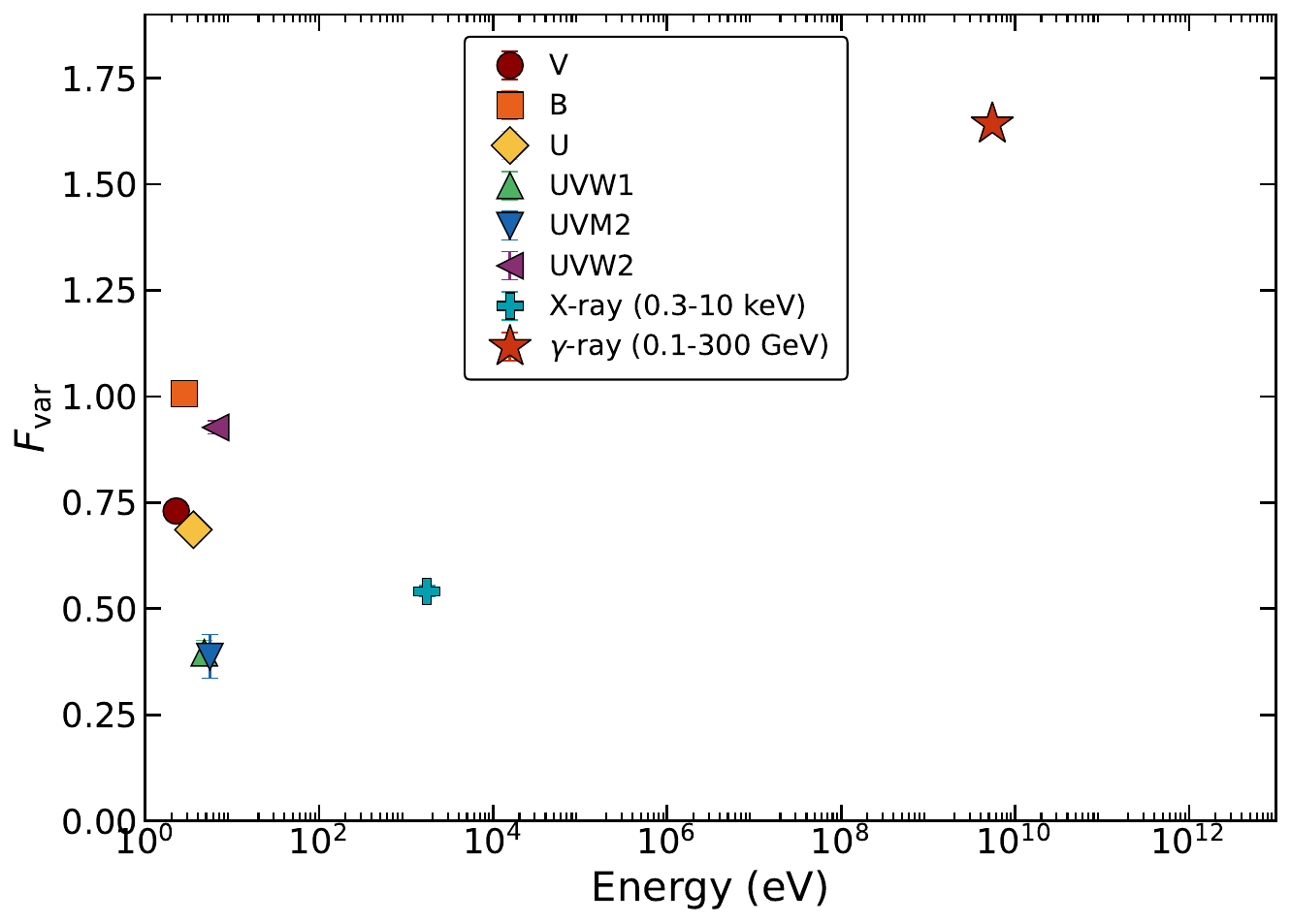}
        \caption{Energy-dependent fractional variability in different energy bands.}
    \label{frac_var}
\end{figure}
where $\rm S^2$ is the variance of the flux, $\rm \overline{F}$ is the mean flux, and $\rm \overline{\sigma^2_{\mathrm{err}}}$ is the mean square of the measurement uncertainties on the flux points. The uncertainty on $F_{\mathrm{var}}$ is also computed following the method provided by \citet{2003MNRAS.345.1271V}\\
\begin{equation}
\rm F_{var,err}=\sqrt{\frac{1}{2N}\left(\frac{\overline{\sigma^{2}_{err}}}{F_{var}\overline{F}^2}\right)^2+\frac{1}{N}\frac{\overline{\sigma^{2}_{err}}}{\overline{F}^2} }
\end{equation}
\vspace{0.5em}
Where, N is the number of simultaneous flux points in the light curve across all energy bands. Figure \ref{frac_var} presents the relationship between $F_{\mathrm{var}}$ and photon energy, while Table~\ref{f_vart} provides a summary of the calculated $F_{\mathrm{var}}$ values for the $\gamma$-ray, X-ray, and UVOT bands. Our analysis indicates that the $F_{\mathrm{var}}$ exhibits a pronounced energy dependence, reflecting the complex and energy stratified emission processes operating in B2\,1420+32. The $\gamma$-ray band (0.1--300~GeV) displays the highest variability ($F_{\rm var}=1.642\pm0.003$), demonstrating that the high-energy emission is extremely sensitive to changes in the relativistic jet. Such large-amplitude variations are naturally expected when the GeV radiation is dominated by EC scattering or when modest fluctuations in the Doppler factor lead to strong amplification of the observed flux. In contrast, the X-ray emission (0.3--10~keV) exhibits moderate variability ($F_{\rm var}=0.541\pm 0.013$), consistent with its origin in SSC and low-energy EC processes.
\begin{table}
\centering
\caption{Fractional amplitude variability ($F_{\rm var}$) of B2 1420+32 in different energy bands with simultaneous data across the light curve.}
\label{f_vart}
\begin{tabular}{lc}
\hline
Energy Band & $F_{\rm var}$ \\
\hline
$\gamma$-ray (0.1--300\,GeV) & $1.642 \pm 0.003$ \\
X-ray (0.3--10\,keV) & $0.541 \pm 0.013$ \\
UVW2 & $0.927 \pm 0.016$ \\
UVM2 & $0.387 \pm 0.052$ \\
UVW1 & $0.396 \pm 0.029$ \\
U & $0.686 \pm 0.028$ \\
B & $1.007 \pm 0.022$ \\
V & $0.730 \pm 0.023$ \\
\hline
\end{tabular}
\end{table}
Significant variability is also observed in the optical and UV bands, with the B and V bands exhibiting particularly large amplitudes ($F_{\rm var}\sim1$). Such pronounced variability suggests that these bands are strongly influenced by the variable non-thermal continuum associated with the jet emission. In the context of a changing-look blazar (CLB) such as B2~1420+32, this behaviour may reflect variations in the synchrotron component, including shifts in the synchrotron peak frequency or changes in the underlying particle energy distribution and spectral shape during different activity states. The UV data further reveal a systematic trend in which $F_{\rm var}$ decreases from the far-UV (UVW2) to the near-UV (UVW1). This behaviour may indicate that the higher-frequency UV bands are more strongly affected by the variable high-energy tail of the synchrotron emission, whereas the lower-frequency UV bands receive an additional contribution from a comparatively less variable emission component.
It is important to note that the fractional variability amplitude ($F_{\rm var}$) can be influenced by observational factors such as temporal sampling, binning,  measurement uncertainties and the number of available observations. \citep{Schleicher2019} quantified these effects for multi-wavelength blazar light curves and found that (i) sparser cadence
alone can shift the recovered $F_{\rm var}$ by up to $\sim$27\%, and (ii) the systematic uncertainty from incomplete sampling increases as
the completeness of the data set decreases. These effects are particularly relevant for our multi-wavelength data set, since Fermi-LAT provides relatively regular monitoring, whereas the Swift-XRT and Swift-UVOT observations are comparatively sparse and irregular; following \citep{Schleicher2019}, the X-ray and optical/UV bands are therefore expected to carry a larger sampling-driven systematic uncertainty than the $\gamma$-ray band. Although we restrict the analysis to simultaneous observations to minimize differences in temporal coverage, this procedure does not completely remove the effects of uneven sampling and different temporal resolutions. The observed $F_{\rm var}$--energy relation should therefore be regarded as an empirical trend rather than a definitive diagnostic of electron cooling or a unique emission mechanism, and should be interpreted alongside the broadband SED results.


To investigate the degree of flux correlation among different energy bands, we
performed a Spearman rank-correlation analysis between the \(\gamma\)-ray,
X-ray, and optical/UV light curves, using the  epochs for which
simultaneous measurements are available in all bands. The resulting correlation
coefficients and null-hypothesis probabilities are summarized in
Table~\ref{tab:spearman}. Strong positive correlations are observed between the
\(\gamma\)-ray and optical/UV bands, with \(\rho\) ranging from 0.74 to 0.88 and
null-hypothesis probabilities below \(10^{-7}\) in all cases, indicating a close
connection between the processes producing the optical/UV and \(\gamma\)-ray
emission. The \(\gamma\)-ray--X-ray correlation is comparatively moderate
(\(\rho = 0.44\), \(P = 1.1\times10^{-3}\)), implying that the X-ray emission
does not track the \(\gamma\)-ray variability as closely as the optical/UV
emission does.
The Spearman analysis above uses contemporaneous measurements and therefore does not by itself establish that the different emission components vary simultaneously. To investigate possible time delays, we applied the Discrete Correlation Function (DCF; \citep{EdelsonKrolik1988}), with flux-randomization/random-subset-selection (FR/RSS; \citealt{1998ApJ...501...82P}) resampling for significance
estimation and the interpolated cross-correlation function (ICCF) as a cross-check, to the $\gamma$-ray light curve against the X-ray and each UVOT light curve. This test requires reasonably dense and contiguous sampling and was therefore restricted to the
two most densely sampled intervals of the light curve: MJD 58850--58901 (Interval S1) and MJD 60047--60088 (Interval S2). These
intervals are defined purely by sampling density. The remaining portions of the light curve do not contain enough simultaneous Swift epochs to constrain a cross-correlation function.

In Interval S1, the $\gamma$-ray/X-ray DCF peaks at a positive lag, with an FR/RSS 68 per cent confidence interval on the centroid lag of [1.4, 8.4] days (median 5.6 days) that excludes zero, indicating that the X-ray emission lags the $\gamma$-ray emission by several days in this interval. In Interval S2, and for all six UVOT bands in both intervals, the FR/RSS confidence intervals on the centroid lag are consistent with zero within the current sampling, and we do not find evidence for a significant lag in these cases. We further tested whether the Interval S1 lag accounts for the moderate $\gamma$-ray/X-ray correlation reported above, by comparing the
zero-lag Spearman coefficient in this interval to the coefficient obtained after shifting the X-ray light curve by the measured lag.
The zero-lag coefficient ($\rho = 0.85$) exceeds the lag-shifted coefficient ($\rho = 0.60$), indicating that the moderate
correlation is not attributable to this lag. Given the sparse and irregular Swift sampling relative to the near-daily Fermi-LAT coverage, these lag constraints should be regarded as indicative rather than definitive, particularly where fewer than approximately 15--20 simultaneous epochs are available. We therefore describe the correlations reported in this section as holding at the sampled epochs, rather than as evidence for a specific causal or strictly simultaneous relationship between the emission components.

\begin{table}
\centering
\caption{Spearman rank correlation coefficients ($\rho$) and corresponding null hypothesis probabilities ($P$-values) between different energy bands.}
\label{tab:spearman}
\begin{tabular}{lcc}
\hline
\textbf{Correlation Pair} & $\boldsymbol{\rho}$ & \textbf{$P$-value} \\
\hline
$\gamma$-ray vs V       & 0.807  & $1.7 \times 10^{-10}$ \\
$\gamma$-ray vs B       & 0.778  & $3.4 \times 10^{-9}$ \\
$\gamma$-ray vs U       & 0.833  & $2.46 \times 10^{-11}$ \\
$\gamma$-ray vs W1      & 0.881  & $6.27 \times 10^{-14}$ \\
$\gamma$-ray vs M2      & 0.744  & $3.64 \times 10^{-8}$ \\
$\gamma$-ray vs W2      & 0.871  & $2.30 \times 10^{-12}$ \\
$\gamma$-ray vs X-ray   & 0.440   & $1.10 \times 10^{-3}$\\
\hline
\end{tabular}
\end{table}

Finally, based on the observed variability and the availability of simultaneous observations across the $\gamma$-ray, X-ray, and optical/UV bands, we identified five distinct flux states, labeled VHE, F1, F2, F3, and F4, which are marked by shaded vertical bands in Figure~\ref{fig:mwl}. The corresponding intervals in the one-day binned \textit{Fermi}-LAT light curve were found to encompass multiple adjacent Bayesian blocks obtained using the algorithm of \citet{Scargle_2013}. For the purpose of spectral analysis, these adjacent blocks were merged within each flux state to improve the photon statistics.
The multi-wavelength data obtained during these states reveal substantial variations in both flux and spectral behaviour. In particular, the VHE period, which coincides with the MAGIC detection of B2~1420+32 at a significance exceeding \(13\sigma\) and a flux level of about 15\% of the Crab Nebula flux above 100 GeV \citep{ATel13412}, provides a particularly important constraint on the high-energy emission. In FSRQs, the GeV $\gamma$-ray emission is commonly interpreted within the framework of EC scattering. However, the detection of VHE photons suggests that the emitting region may lie beyond the BLR, where internal \(\gamma\gamma\) absorption is reduced and infrared photons from the dusty torus can act as seed photons for the EC process. At the same time, the pronounced optical enhancement observed during the high-activity states indicates that synchrotron photons produced within the jet may also provide seed photons for SSC emission. This possibility is particularly relevant for a changing-look blazar such as B2~1420+32, which may display BL~Lac-like behaviour during some intervals.
These considerations indicate that temporal variability alone is insufficient to uniquely identify the dominant radiative mechanism. Therefore, to investigate the origin of the emission in the selected flux states and to constrain the underlying physical parameters of the source, we perform a detailed SED analysis in Section~\ref{broadband}.

\section{BROADBAND SPECTRAL AND HIGH-ENERGY EMISSION ANALYSIS:}\label{broadband}
\indent

Motivated by the variability properties described above, we next investigate the broadband spectral behaviour of B2~1420+32 in the selected activity states. We performed a detailed spectral analysis of the five flux states--VHE, F1, F2, F3, and F4--identified from the temporal analysis. For each state, simultaneous multi-wavelength observations were used to construct the broadband spectrum. This state-wise approach enables a systematic comparison of broadband emission properties and provides insight into the evolution of spectral parameters of the particle distribution across different activity states.
 The VHE flux state is defined over the period MJD,58868--58872, during which the source exhibited enhanced very-high-energy (VHE) $\gamma$-ray emission. Within this interval, the MAGIC telescopes observed B2~1420+32 during the nights MJD,58868.3--58870.3, and the detailed analysis of these VHE observations is presented by Acciari et al. (2021). The corresponding VHE spectrum is shown in Fig.~4 of Acciari et al. (2021). To incorporate the spectral information from the MAGIC observations into our broadband analysis, we digitised the published EBL-corrected VHE spectrum and extracted the individual spectral points together with their associated uncertainties. The EBL correction adopted in Acciari et al. (2021) follows the model of Dom'inguez et al. (2011). These de-absorbed VHE spectral points were then included in the broadband SED fitting together with the contemporaneous \textit{Fermi}-LAT, \textit{Swift}-XRT, and \textit{Swift}-UVOT data.
For each flux state, we examined the spectral behaviour in the $\gamma$-ray and X-ray regimes. In the $\gamma$-ray band, the spectra were fitted with a power-law (PL) and a log-parabola (LP) model in order to characterize the spectral shape and identify possible curvature associated with flux changes. The LP model is expressed as
\begin{equation}
\frac{dN}{dE}
=
N_{0}
\left(\frac{E}{E_{0}}\right)^{-\alpha-\beta \log(E/E_{0})},
\end{equation}
where $N_{0}$ denotes the normalization, $\alpha$ is the photon index at the pivot energy $E_{0}$, fixed at 917.08 MeV, and $\beta$ quantifies the spectral curvature. The PL model is given by
\begin{equation}
\frac{dN}{dE}
=
N_{0}
\left(\frac{E}{E_{0}}\right)^{-\Gamma},
\end{equation}
where $N_{0}$ is the normalization and $\Gamma$ is the photon index. These empirical models are commonly used in blazar studies to describe curved as well as hardening or softening $\gamma$-ray spectra \citep{Massaro2004}.
Using these models, we quantified the presence of spectral curvature in the $\gamma$-ray spectra of the selected flux states by performing a likelihood-ratio test using the statistic \citep{Nolan2012}
\begin{equation}
TS_{\rm curve} = 2\left[\log \mathcal{L}(\mathrm{LP}) - \log \mathcal{L}(\mathrm{PL})\right].
\end{equation}

Following \citet{Nolan2012}, a value of \(TS_{\rm curve} > 16\) is
taken to indicate statistically significant curvature. The resulting
\(TS_{\rm curve}\) values for all flux states are listed in
Table~\ref{TStable}. Significant curvature is found for the F1 and VHE states,
whose \(\gamma\)-ray spectra are accordingly described by a log-parabola model,
while a power law is sufficient for the remaining states. The F4 state, with
\(TS_{\rm curve} = 14.09\), lies just below the adopted threshold and is
therefore best regarded as a marginal case rather than a clear non-detection of
curvature. Across the five states the measured \(TS_{\rm curve}\) follows the same ordering
as the overall detection significance, increasing monotonically from
\(TS_{\rm curve} = 2.49\) at \(TS = 134\) in the F2 state to
\(TS_{\rm curve} = 45.35\) at \(TS = 10214\) in the F1 state. This indicates
that the ability to detect curvature in these spectra is governed principally by
the available photon statistics, and that the absence of significant curvature
in the fainter states need not imply an intrinsically different spectral shape.
A physical evolution of the curvature with activity level cannot be excluded,
but the present data do not separate it from the statistical trend.

\FloatBarrier

\begin{table*}
\centering
\small
\renewcommand{\arraystretch}{1.3}
\setlength{\tabcolsep}{7pt}
\caption{Likelihood-fit spectral parameters for B2~1420+32 obtained by fitting the $\gamma$-ray spectrum with power-law (PL) and log-parabola (LP) models across different flux states during MJD 58818--60721.}
\label{TStable}
\begin{tabular}{lccccc}
\hline
State & Period & Model & TS & $-\log \mathcal{L}$ & $TS_{\rm curve}$ \\
\hline

F1 State & 58845--58911 & log-parabola & 10214.09 & 19146.88 & 45.35 \\
         &              & power-law    & 10043.39 & 19169.56 & -- \\

VHE State & 58868--58872 & log-parabola & 3417.77 & 5801.30 & 19.93 \\
          &              & power-law    & 3291.07 & 5811.26 & -- \\

F2 State & 59437--59461 & log-parabola & 133.60 & 6048.06 & 2.49 \\
         &              & power-law    & 131.19 & 6049.30 & -- \\

F3 State & 60045--60100 & log-parabola & 1051.89 & 11571.48 & 5.76 \\
         &              & power-law    & 1040.56 & 11574.37 & -- \\

F4 State & 60484--60530 & log-parabola & 1684.60 & 12880.40 & 14.09 \\
         &              & power-law    & 1667.56 & 12887.45 & -- \\
\hline
\end{tabular}
\end{table*}
Swift observations obtained during the period MJD~58818--60721 were used to generate the X-ray spectra. For each selected flux state, the cleaned event files corresponding to the individual observation IDs within that interval were combined to obtain a joint spectrum. The \texttt{xselect} tool was then used to extract the source and background spectra for each flux state. The ancillary response files (ARFs) were produced with \texttt{xrtmkarf}, and the \texttt{grppha} task was used to group the spectra, ensuring a minimum of 20 counts per bin. The resulting spectra were analyzed with \texttt{XSPEC} using the models \texttt{Tbabs$\times$powerlaw}, \texttt{Tbabs$\times$logparabola}, and \texttt{Tbabs$\times$bknpower}. Galactic absorption was accounted for by fixing the hydrogen column density (nH) was fixed at $1.07\times10^{20}~{\rm cm^{-2}}$ \citep{HI4PI2016}.

While all three models yielded statistically acceptable fits, the log-parabola
model provided a better description of the spectra than either the
power-law or the broken power-law, suggesting the presence of intrinsic curvature in the X-ray band \citep{Massaro2004}. The best-fit parameters obtained from the log-parabola fits are summarized in Table~\ref{tab:tbabs_model_logparabola}. In some flux states, the curvature parameter \(\beta\) is found to be negative, indicating a concave spectral shape in which the spectrum hardens with energy. 

For the Swift-UVOT analysis, images from individual filters corresponding to each selected flux state were combined using \texttt{uvotimsum}, and the fluxes were extracted with \texttt{uvotproduct}. The optical/UV spectra could not be adequately described by a simple power-law model, suggesting the possible presence of additional emission components besides the jet emission. Since the optical/UV fluxes possess very small statistical uncertainties and can disproportionately influence the broadband spectral fitting, additional systematic uncertainties were introduced to balance the relative contributions from different energy bands and to obtain an acceptable fit with a reduced $\chi^{2} \sim 1$. The corrected X-ray, optical/UV, and $\gamma$-ray fluxes were subsequently converted into Pulse Height Analyser (PHA) files using the HEASARC utility \texttt{ftflx2xsp} for broadband spectral modelling.

We modeled the broadband SEDs corresponding to different flux states using a one-zone leptonic emission framework \citep{2019MNRAS.484.3168S, 2021MNRAS.504..416S}. In this model, the radiative output is assumed to originate from a homogeneous spherical region (or blob) of radius \(R\), moving relativistically along the jet with a bulk Lorentz factor \(\Gamma_{\rm b}\) at a small angle \(\theta\) to the observer's line of sight. Such relativistic motion, combined with the narrow viewing angle, results in Doppler boosting of the observed flux, characterized by the Doppler factor \(\delta = [\Gamma_{\rm b}(1 - \beta \cos\theta)]^{-1}\), where \(\beta\) is the velocity of the emitting region in units of the speed of light. We assumed the emission zone is populated by non-thermal electron with distribution $n(\gamma)$ . These relativistic electrons radiate via synchrotron emission and IC scattering. B2 1420+32 is formally classified as a FSRQ based on the presence of strong broad emission lines and a luminous accretion disk features. However, the source has shown episodes where these broad emission lines become extremely weak or disappear, causing it to appear as a BL Lac object. During these BL Lac like states, the broadband SED can be adequately described using only synchrotron and SSC processes, with negligible contribution from external photon fields.  In this scenario, the synchrotron photons produced within the jet act as seed photons for IC scattering, making  SSC the dominant mechanism responsible for the high-energy emission.

To compute the model spectra within the one-zone leptonic framework involving
synchrotron and IC emission, we express the electron Lorentz factor \(\gamma\)
in terms of a new variable \(\xi\) such that \(\xi = \gamma\sqrt{\mathbb{C}}\),
where \(\mathbb{C} = 1.36\times10^{-11}\,\delta B/(1+z)\)
\citep{2025JHEAp..45..438A}. Following \citet{1984RvMP...56..255B}, the synchrotron flux at photon energy \(\epsilon\) can then be obtained as:

\begin{equation}\label{eq:syn_flux}
F_{\rm syn}(\epsilon) = \frac{\delta^3(1+z)}{d_L^2} V \mathbb{A} \int_{\xi_{\min}}^{\xi_{\max}} f(\epsilon/\xi^2)\, n(\xi)\, d\xi,
\end{equation}

where \(d_L\) is the luminosity distance, \(V\) is the volume of the emission region, and \(\mathbb{A} = \frac{\sqrt{3}\pi e^3 B}{16 m_e c^2 \sqrt{\mathbb{C}}}\). The limits \(\xi_{\min}\) and \(\xi_{\max}\) correspond to the minimum and maximum electron energies, respectively, and \(f(x)\) denotes the synchrotron emissivity function \citep{1986rpa..book.....R}.  

Similarly, the observed SSC flux at photon energy \(\epsilon\) is obtained as:

\begin{equation}\label{eq:ssc_flux}
\begin{split}
F_{\rm ssc}(\epsilon) = \frac{\delta^3(1+z)}{d_L^2} V \mathbb{B} \epsilon &
\int_{\xi_{\min}}^{\xi_{\max}} \frac{1}{\xi^2} 
\int_{x_1}^{x_2} \frac{I_{\rm syn}(\epsilon_i)}{\epsilon_i^2} \\
& \times f(\epsilon_i, \epsilon, \xi/\sqrt{\mathbb{C}})\, d\epsilon_i\, n(\xi)\, d\xi,
\end{split}
\end{equation}

where \(\epsilon_i\) is the energy of the incident photon, \(\mathbb{B} = \frac{3}{4}\sigma_T\sqrt{\mathbb{C}}\), \(I_{\rm syn}(\epsilon_i)\) is the synchrotron intensity, and the limits of integration are defined as \(\rm x_1 = \frac{\mathbb{C} \epsilon}{4\xi^2(1-\sqrt{\mathbb{C}}\epsilon/\xi m_e c^2)}\) and \(\rm x_2 = \frac{\epsilon}{(1-\sqrt{\mathbb{C}}\epsilon/\xi m_e c^2)}\). The function \(f(\epsilon_i, \epsilon, \xi)\) describes the IC scattering kernel and is expressed as:

\begin{equation}
f(\epsilon_i, \epsilon, \xi) =
2q\ln q + (1 + 2q)(1 - q) + \frac{\kappa^2 q^2 (1 - q)}{2(1 + \kappa q)},
\end{equation}

where \(\rm q = \frac{\mathbb{C}\epsilon}{4\xi^2\epsilon_i(1 - \sqrt{\mathbb{C}}\epsilon/\xi m_e c^2)}\) and \(\rm \kappa = \frac{4\xi\epsilon_i}{\sqrt{\mathbb{C}} m_e c^2}\).  

\vspace{0.30cm}

Similarly, the observed external Compton (EC) flux at photon energy $\epsilon$ can be written as:

\begin{equation}\label{eq:ec_flux}
\begin{split}
F_{\rm ec}(\epsilon) =
\frac{\delta^{3}(1+z)}{d_{L}^{2}}
\, V \mathbb{B} \epsilon
\int_{\xi_{\min}}^{\xi_{\max}}
\frac{1}{\xi^{2}}
\int_{\epsilon^{*}_{1}}^{\epsilon^{*}_{2}}
\frac{I^{*}(\epsilon^{*})}{\epsilon^{*2}}\\
\times
\,f\left(\epsilon^{*}, \epsilon, \frac{\xi}{\sqrt{\mathbb{C}}}\right)
\, d\epsilon^{*}
\, n(\xi)
\, d\xi ,
\end{split}
\end{equation}


\begin{table*}
\centering
\renewcommand{\arraystretch}{1.6}
\setlength{\tabcolsep}{8.5pt}

\caption{Best-fit X-ray spectral parameters for B2~1420+32 in different flux states using the log-parabola model.}
\label{tab:tbabs_model_logparabola}

\begin{tabular}{lcccc}
\hline
State & $N~(\times10^{-4})$ & $\alpha$ & $\beta$ & $\chi^{2}/\mathrm{dof}$ \\
\hline
F1  & $4.94_{-1.8}^{+1.8}$ & $1.78_{-0.05}^{+0.05}$ & $-0.31_{-0.10}^{+0.10}$ & $165.07/126$ \\
VHE & $7.97_{-4.2}^{+4.3}$ & $1.97_{-0.01}^{+0.01}$ & $-0.23_{-0.02}^{+0.02}$ & $48.80/32$ \\
F2  & $1.49_{-0.2}^{+0.2}$ & $1.50_{-0.30}^{+0.23}$ & $-0.38_{-0.05}^{+0.06}$ & $3.44/6$ \\
F3  & $3.22_{-1.6}^{+1.6}$ & $1.56_{-0.09}^{+0.08}$ & $0.03_{-0.05}^{+0.06}$ & $67.10/70$ \\
F4  & $8.00_{-3.9}^{+4.0}$ & $1.54_{-0.01}^{+0.07}$ & $-0.16_{-0.01}^{+0.01}$ & $67.81/76$ \\
\hline
\end{tabular}

\end{table*}

Equations~\ref{eq:syn_flux}, \ref{eq:ssc_flux}, and~\ref{eq:ec_flux} were solved numerically, and the resulting code was implemented as a local convolution model within XSPEC for the statistical fitting of broadband SEDs. This convolution framework provides flexibility in modeling spectra for any arbitrary electron energy distribution \(n(\xi)\), allowing a consistent description of the observed emission across the entire energy range. In this work, we assumed a broken power-law energy distribution:

\begin{equation}
n(\xi)d\xi =
\begin{cases}
K \xi^{-p} d\xi, & \xi_{\min} < \xi < \xi_b, \\
K \xi_b^{q-p} \xi^{-q} d\xi, & \xi_b < \xi < \xi_{\max},
\end{cases}
\quad \mathrm{cm^{-3}},
\label{equation(6)}
\end{equation}
where \(\xi_b\) corresponds to the break energy, \(K\) is the normalization constant, and \(p\) and \(q\) represent the spectral indices below and above the break, respectively. 
 Within this model, the broad emission involving synchrotron, SSC and EC emission is primarily governed by  key parameters: \(\xi_{\rm b}\), \(\xi_{\min}\), \(\xi_{\max}\), \(p\), \(q\), \(B\), \(R\), \(\Gamma_{\rm b}\), \(\theta\), temperature \(T\) and the normalization \(N\). The code also allows the inclusion of the jet power (\(P_{\rm jet}\)) as a free parameter; however, in such cases, the normalization \(N\) must be fixed to maintain consistency.  
To avoid over-parameterization and ensure physical interpretability, we adopted a minimalistic approach by keeping only a subset of parameters free during the fitting procedure. Specifically, \(p\), \(q\), \(\Gamma_{\rm b}\), \(B\) and  \(\xi_{\rm b}\) were treated as free parameters, while the remaining quantities were fixed to their typical values inferred from the broadband spectrum. The parameters were fixed due to the limited spectral coverage available across the optical/UV, X-ray, and $\gamma$-ray energy ranges. The jet viewing angle was fixed at \(\theta = 2^{\circ}\). 
The kinetic power of the jet was estimated using the relation:  
\[
P_{\rm jet} = \pi R^{2} \Gamma_{\rm b}^{2} \beta c \, (U_{\rm e} + U_{\rm p} + U_{\rm B}),
\]
where \(U_{\rm e}\), \(U_{\rm p}\), and \(U_{\rm B}\) represent the energy densities of relativistic electrons, cold protons, and the magnetic field, respectively \citep{2008MNRAS.385..283C}. In this framework, the protons are assumed to be cold and do not contribute to the radiative processes, consistent with a leptonic emission scenario. Furthermore, the number of protons is taken to be equal to the number of non-thermal electrons, implying a heavy jet composition. 

\begin{figure*}
\centering
\hspace{-0.07\linewidth}
\begin{subfigure}[b]{0.40\linewidth}
\centering
\includegraphics[width=0.9\linewidth,angle=-90,origin=c]{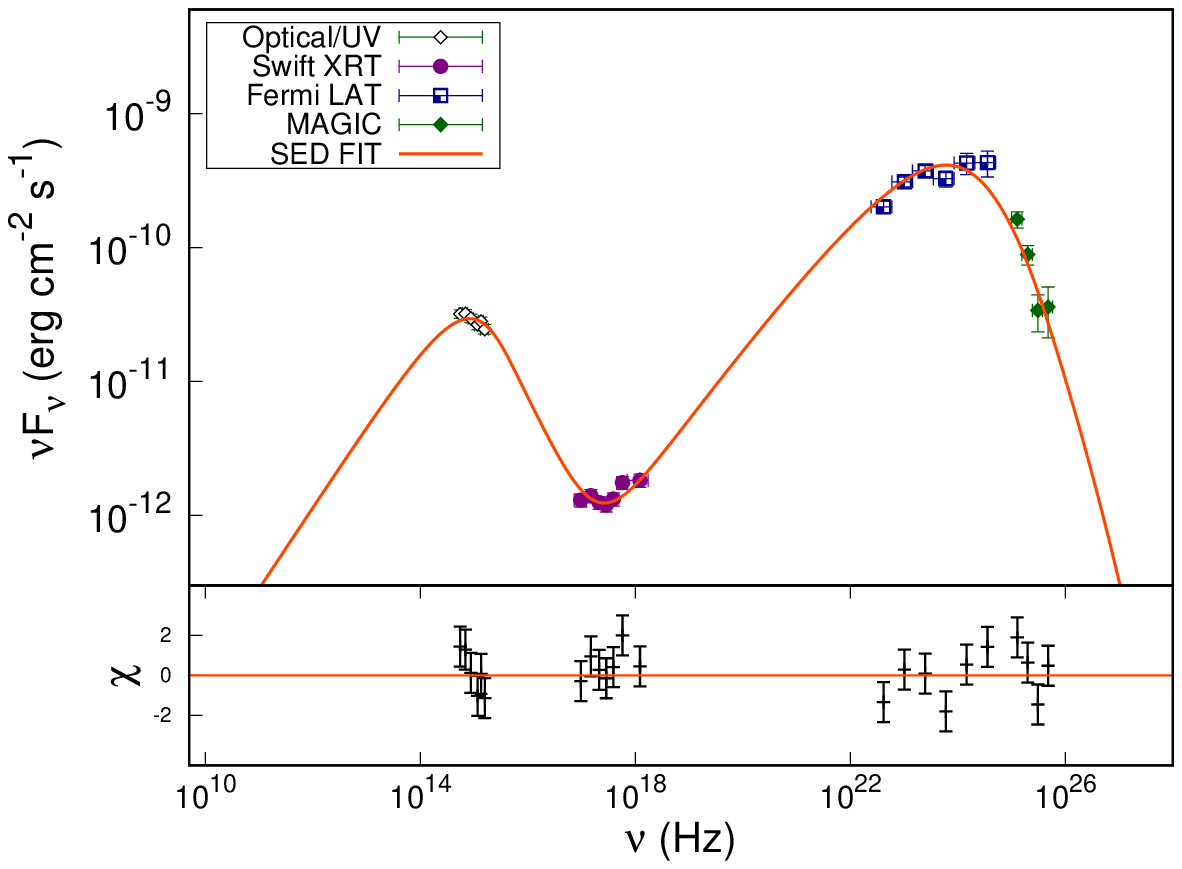}
\end{subfigure}
\hspace{0.1\linewidth}
\begin{subfigure}[b]{0.40\linewidth}
\centering
\includegraphics[width=0.9\linewidth,angle=-90,origin=c]{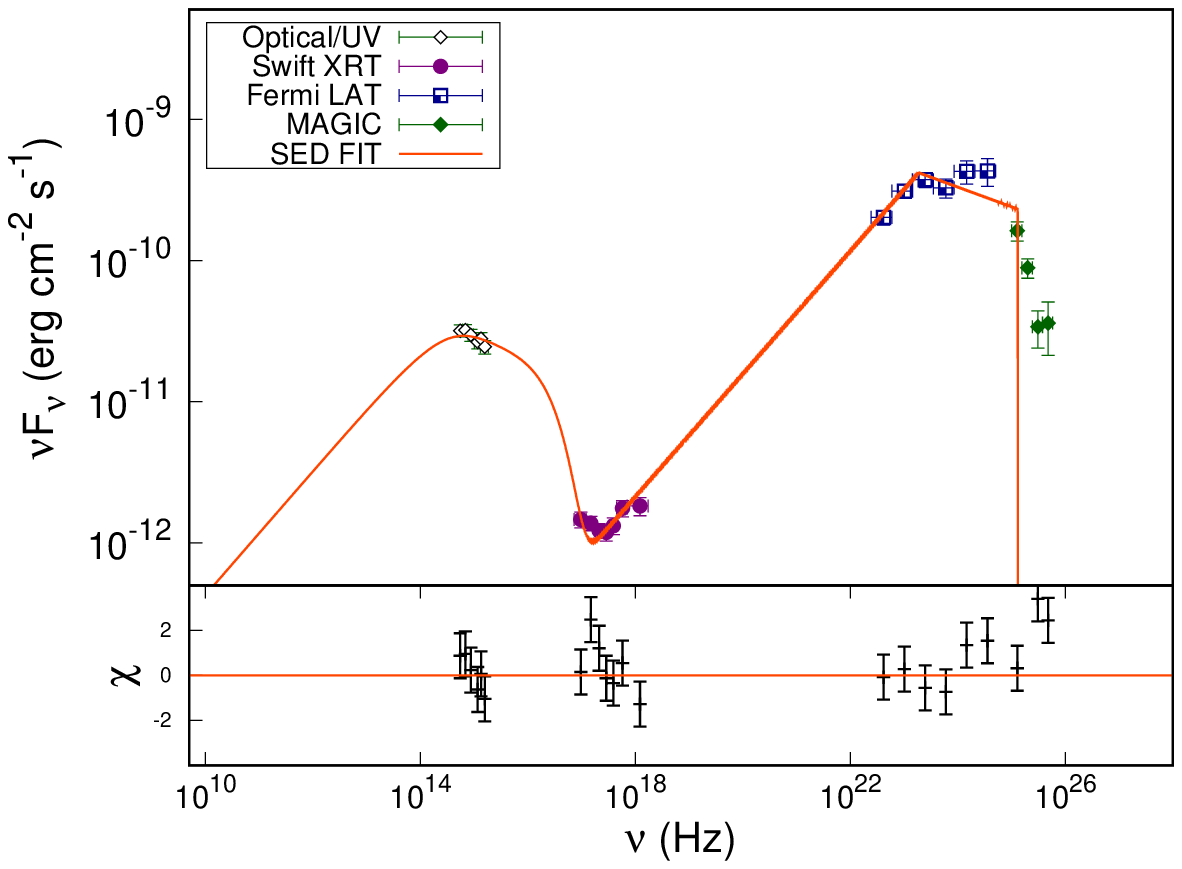}
\end{subfigure}

\caption{Broadband SEDs of B2 1420+32 during the VHE state (MJD 58868--58872). The solid red curve represents the model fit. The left panel shows the SED considering only the SSC process, while the right panel corresponds to the SED considering only the EC process.}
\label{fig:sed_1states}

\end{figure*}

\begin{table*}
\centering
\caption{Best-fit model parameters for B2~1420+32 obtained from fitting the broadband SED model to the VHE, F1, F2, F3, and F4 states assuming an SSC-only scenario. The table is organized into four sections. The top section lists the free parameters varied during the fitting procedure: the bulk Lorentz factor of the emitting region ($\Gamma_{b}$), the magnetic field strength ($B$) in units of G, the broken power-law indices $p$ and $q$, and the break energy parameter $\xi_{b}$. The upper-middle section includes the fixed parameters: $\xi_{\min}$ (in units of $10^{-6}\sqrt{\rm keV}$) and $\xi_{\max}$ (in units of $\sqrt{\rm keV}$), corresponding to the minimum and maximum electron energies, respectively. The emission region radius ($R=10^{16}$\,cm) and the jet inclination angle ($\theta = 2^\circ$) were kept fixed throughout the fitting procedure. The lower-middle section includes the converted parameters $\gamma_{\min}$, $\gamma_{b}$, and $\gamma_{\max}$ (in units of $10^{7}$), derived from $\xi$. The bottom section lists $\log P_{\rm jet}$ (in units of erg\,s$^{-1}$) and $\chi^{2}/\mathrm{dof}$.}
\label{tab:ssc_only_combined}

\setlength{\tabcolsep}{7pt}
\renewcommand{\arraystretch}{1.5}

\begin{tabular}{lccccc}
\hline
\textbf{Parameter} & \textbf{VHE} & \textbf{F1} & \textbf{F2} & \textbf{F3} & \textbf{F4} \\
\hline

\multicolumn{6}{l}{\textit{Free parameters}} \\
$\Gamma_{b}$ & $32.65_{-3.58}^{+6.13}$ & $28.62_{-0.99}^{+0.98}$ & $13.50_{-0.92}^{+1.13}$ & $20.03_{-1.88}^{+3.65}$ & $28.90_{-0.98}^{+2.03}$ \\
$B$ & $0.020_{-0.001}^{+0.006}$ & $0.020_{-0.001}^{+0.002}$ & $0.031_{-0.002}^{+0.003}$ & $0.034_{-0.002}^{+0.001}$ & $0.021_{-0.003}^{+0.008}$ \\
$p$ & $1.78_{-0.01}^{+0.03}$ & $2.06_{-0.03}^{+0.03}$ & $3.17_{-0.06}^{+0.05}$ & $2.50_{-0.04}^{+0.02}$ & $2.10_{-0.01}^{+0.02}$ \\
$q$ & $4.65_{-0.08}^{+0.17}$ & $4.90_{-0.14}^{+0.16}$ & $5.21_{-0.47}^{+1.60}$ & $5.52_{-0.01}^{+0.01}$ & $ 5.05_{--}^{--}$ \\
$\xi_{b}$ & $0.060_{-0.002}^{+0.075}$ & $0.063_{-0.001}^{+0.001}$ & $0.142_{-0.005}^{+0.110}$ & $0.055_{-0.005}^{+0.052}$  & $0.051_{-0.002}^{+0.011}$ \\

\hline
\multicolumn{6}{l}{\textit{Fixed parameters}} \\
$\xi_{\min}$ & 1.00 & 531 & 2372 & 1.00 & 506 \\
$\xi_{\max}$ & 25 & 25 & 25 & 25 & 25 \\

\hline
\multicolumn{6}{l}{\textit{Converted parameters}} \\
$\gamma_{\min}$ & 0.466 & 246.68 & 1008.27 & 0.368 & 229.40 \\
$\gamma_{\rm b}$ & 28458.72 & 29267.04 & 60359.90 & 20222.36 & 23121.80 \\
$\gamma_{\max}$ & 1.17 & 1.16 & 1.06 & 0.92 & 1.13 \\
\hline
$\log P_{\rm jet}$ & 48.68 & 46.56 & 45.71 & 45.62 & 46.72  \\
$\chi^{2}/{\rm dof}$ & 69.79/46 & 188.13/137 & 15.30/15 & 80.45/80 & 106.77/89 \\
\hline
\end{tabular}

\end{table*}

\begin{table*}
\centering
\caption{Best-fit model parameters for B2~1420+32 obtained by fitting the broadband SED model to the VHE, F1, F2, F3, and F4 states considering only the EC process. The table is organized into three sections. The top section lists the free parameters varied during the fit: the bulk Lorentz factor of the electron energy distribution ($\Gamma_b$); the magnetic field strength ($B$), expressed in units of G; the broken power-law indices $p$ and $q$; and $\xi_b$. The middle section includes the converted parameters $\gamma_{\min}$, $\gamma_b$, and ($\gamma_{\max}$ in units of $10^{4}$), derived from $\xi$. The bottom section lists $\log P_{\rm jet}$ (in units of erg\,s$^{-1}$) and $\chi^2/\mathrm{dof}$. The emission region radius  was fixed at ($10^{17}$ cm) and jet inclination angle ($\theta = 2^\circ$) were fixed. The target photon temperature was fixed in the range 800--1000~K. The parameters $\xi_{\min}$ and $\xi_{\max}$ are fixed to $1 \times 10^{-6}$ and 1, respectively, for all states. The parameters $\xi_b$, $\xi_{\min}$, and $\xi_{\max}$ represent the break, minimum, and maximum electron energies, respectively, with all energies expressed in units of $\sqrt{\mathrm{keV}}$. Uncertainties on the VHE-state parameters and $\log P_{\rm jet}$ are not reported: the fit did not converge to a stable error estimate for this state. The best-fit central values are retained for comparison, but should be treated as less robust than the fully converged fits obtained for the other four states.}
\label{tab:parameters_EC_combined}

\setlength{\tabcolsep}{7pt}
\renewcommand{\arraystretch}{1.5}

\begin{tabular}{lccccc}
\hline
\textbf{Parameter} & \textbf{VHE} & \textbf{F1} & \textbf{F2} & \textbf{F3} & \textbf{F4} \\
\hline

\multicolumn{6}{l}{\textit{Free parameters}} \\
$\Gamma_{b}$ & $47.801^{\ast}$ & $45.12_{--}^{--}$ & $29.07_{-0.01}^{+0.01}$ & $35.47_{--}^{--}$ & $46.36_{-9.90}^{+11.00}$ \\
$B$ & $0.676^{\ast}$ & $0.880_{-0.071}^{+0.140}$ & $1.560_{-0.507}^{+0.021}$ & $1.111_{-0.107}^{+0.090}$ & $0.911_{-0.101}^{+0.103}$ \\
$p$ & $2.12^{\ast}$ & $2.12_{-0.02}^{+0.01}$ & $2.02_{--}^{--}$ & $2.10_{-0.01}^{+0.01}$ & $2.07_{-0.03}^{+0.06}$ \\
$q$ & $3.31^{\ast}$ & $3.04_{-0.13}^{+0.16}$ & $3.42_{-0.23}^{+0.11}$ & $3.51_{-0.01}^{+0.11}$ & $ 3.25_{-0.06}^{+0.13}$ \\
$\xi_{b}$ & $0.038^{\ast}$ & $0.025_{-0.001}^{+0.001}$ & $0.007_{-0.002}^{+0.001}$ & $0.013_{-0.001}^{+0.001}$  & $0.004_{-0.010}^{+0.010}$ \\

\hline

\multicolumn{6}{l}{\textit{Converted parameters}} \\
$\gamma_{\min}$ & 0.085 & 0.074 & 0.053 & 0.063 & 0.073 \\
$\gamma_{\rm b}$ & 3233.37 & 1840.44 & 368.22 & 819.50 & 291.13 \\
$\gamma_{\max}$ & 8.50 & 7.36 & 5.26 & 6.30 & 7.28 \\

\hline
$\log P_{\rm jet}$ & $-{}^{\ast}$ & 46.38 & 45.30 & 45.60 & 46.54  \\
$\chi^{2}/{\rm dof}$ & 119.76/46 & 197.48/137 & 16.6/15 & 81.10/80 & 97.97/89 \\
\hline
\end{tabular}
\vspace{2pt}
\newline
{\footnotesize $^{\ast}$Fit did not converge to a stable uncertainty estimate for this state; central value reported only.}

\end{table*}

Given the changing-look nature of B2~1420+32, the broadband SED may reflect contributions from different radiative components depending on the source state. The low-energy component, extending from optical to X-ray frequencies, is naturally attributed to synchrotron radiation from relativistic electrons in a magnetized emission region. In contrast, the high-energy emission may arise through different leptonic processes: in a BL~Lac-like state, the X-ray and $\gamma$-ray emission may be dominated by the SSC mechanism, whereas in a more FSRQ-like state, both SSC and EC processes may contribute significantly. The primary objective of the present modeling is therefore to determine which of these radiative scenarios best explains the observed X-ray and $\gamma$-ray emission across the selected flux states. We begin by examining whether the broadband SED can be reproduced within a pure SSC framework.

\subsection{Considering only SSC}
\label{ssc_only}
We first examine whether the broadband SED of B2~1420+32 can be described within a homogeneous one-zone synchrotron and SSC scenario. We find that the synchrotron and SSC components can reproduce the broadband spectra across all selected flux states, yielding statistically acceptable fits. The best-fitting model parameters, together with the corresponding reduced $\chi^{2}$ values and the fixed parameter values obtained in the modeling, are listed in Table~\ref{tab:ssc_only_combined}. For illustration, the SED corresponding to the VHE state is shown in Figure~\ref{fig:sed_1states}. In particular, the SSC model is able to reproduce the observed MAGIC spectral measurements for the VHE-flare interval.

However, statistical agreement alone does not guarantee physical plausibility. To asses the physical consistency of the model, we convert the electron energy parameter $\xi$, into the corresponding electron Lorentz factor $\gamma$ through the transformation 
\begin{equation}
\gamma=\frac{\xi}{M}
\end{equation}
where
\begin{equation}
M = 1.36 \times 10^{-11}
\frac{B}{(1+z)\Gamma}
\left(
1 - \sqrt{1 - \frac{1}{\Gamma^2}} \cos\theta
\right).
\end{equation}
%

This relation incorporates Doppler boosting and cosmological effects, and links the numerical model parameter $\xi$ to the electron Lorentz factor $\gamma$ in the comoving frame.
Applying this transformation, we find that the derived values of $\gamma_{\min}$ span a wide range across the different flux states, ranging from values below unity to approximately $10^{3}$. Values of $\gamma_{\min}<1$ are unphysical, since the electron Lorentz factor is required to satisfy $\gamma \geq 1$. The large state-to-state variation further suggests that $\gamma_{\min}$ is only weakly constrained within the SSC-only scenario. This behaviour indicates that, in a pure SSC framework, $\gamma_{\min}$ is likely compensating for model degeneracies rather than tracing genuine changes in the low-energy cutoff of the electron distribution.
Similarly, the inferred break Lorentz factor, $\gamma_{b}$, lies in the range of $\sim (2$--$6)\times10^{4}$. These values are substantially larger than those typically inferred in one-zone leptonic models of powerful FSRQ-like blazars, where $\gamma_{b}$ is more commonly of order $10^{3}$. Such large $\gamma_{b}$ values would require unusually efficient particle acceleration and/or comparatively weak radiative cooling in a pure SSC framework.
Taken together, these results show that the SSC-only model, although statistically acceptable, requires physically disfavoured values of $\gamma_{\min}$ and $\gamma_{b}$. 
An additional difficulty with the SSC-only interpretation is the large departure from equipartition implied by the model parameters. The ratio of electron to magnetic energy density, $U_{e}/U_{B}$, ranges from $\sim54$ in the F2 state to values of $\sim 217$--$780$ in the remaining flux states, indicating that the emitting region is strongly particle dominated. Such large departures from equipartition are generally considered energetically unfavourable, as they require a substantial excess of energy to be stored in relativistic particles compared to the magnetic field. The energetic requirements of the SSC-only description are also difficult to
sustain. Reproducing the VHE-state spectrum requires
\(\log(P_{\rm jet}/{\rm erg\,s^{-1}}) = 48.68\), some two orders of magnitude
above the accretion-disk luminosity of \(\sim 2\times10^{46}\)
erg\,s\(^{-1}\) inferred for this source from its broad-line emission
\citep{2021A&A...647A.163M}, and super-Eddington for any black hole mass below
\(\sim 1.6\times10^{10}\,M_{\odot}\). The corresponding SSC+EC fit requires
\(\log(P_{\rm jet}/{\rm erg\,s^{-1}}) = 46.79\), a factor of about 78 lower and
consistent with values typical of powerful FSRQ jets. These findings suggest that a pure SSC scenario is unlikely to provide a physically realistic description of the broadband emission and instead favour the presence of an additional external photon field contributing to the observed $\gamma$-ray emission through EC scattering.

\subsection{Considering Only EC}
\label{Ec_only}
We next examine whether the broadband SED of B2~1420+32 can be reproduced within a pure EC framework. In this scenario, the low-energy optical/UV emission is attributed to synchrotron radiation from relativistic electrons in the jet, while the X-ray and $\gamma$-ray emission are produced through inverse Compton scattering of external photon fields by the same electron population. The external photon field is assumed to originate either from the infrared radiation of the dusty torus or from the BLR, and is approximated by a blackbody distribution with characteristic temperatures of $\sim 800-1000$ K and $\sim4.2\times10^{4}$ K, respectively.

The EC-only model again provides acceptable fits to the optical--GeV data. The best-fitting EC model parameters, together with the corresponding reduced $\chi^{2}$ values, are listed in Table~\ref{tab:parameters_EC_combined}. For illustration, the SED corresponding to the VHE state is shown in Figure~\ref{fig:sed_1states}. We note that the EC-only fit to the VHE state did not converge to well-constrained parameter uncertainties (Table~\ref{tab:parameters_EC_combined}), consistent with the difficulty this scenario has in reproducing the observed VHE emission once the seed-photon field is limited by Klein--Nishina suppression, as discussed below.

In these fits, the break Lorentz factor $\gamma_{b}$ lies in the range of $\sim 3\times10^{2}$ to $\sim 3.2\times10^{3}$, which is broadly consistent with values expected from radiative cooling in the presence of strong external photon fields.
However, the derived values of $\gamma_{\min}$ remain sub-relativistic ($\sim 0.05$--$0.09$) across all flux states. Such values are not physically acceptable, as they correspond to electron energies below the rest-mass energy. This indicates that the low-energy end of the electron distribution is not well constrained in the EC-only scenario. The inferred equipartition ratios, $U_{e}/U_{B}$, are $\sim492$, $\sim311$, $\sim171$, $\sim46$, and $\sim10$ for the VHE, F1, F2, F3, and F4 states,
respectively. These values indicate that the emitting region remains particle dominated in all flux states, although the departure from equipartition is smaller than in the SSC-only scenario. The large electron energy densities required in the VHE, F1, and F2 states suggest that the EC-only model still struggles to achieve energetically favourable solutions.

Another limitation is that, although the EC component reproduces the GeV emission satisfactorily, it significantly underpredicts the observed VHE flux and fails to account for the source's emission into the VHE regime. The predicted VHE emission remains well below the MAGIC-measured spectrum for the VHE-flare interval, showing that the EC-only configuration does not provide sufficient high-energy output under physically reasonable parameter choices.
Taken together, these results show that, although the EC-only model provides acceptable description of the optical--GeV emission, it cannot reproduce the VHE component due to extreme  Klein--Nishina cut-off and leaves the low-energy electron parameters poorly constrained. These limitations indicate that a pure EC scenario is not sufficient to explain the full broadband emission, and that additional radiative contributions are required.

\subsection{Combined SSC+EC Scenario}
\label{com}
Finally, we model the broadband SED of B2~1420+32 using a
synchrotron+SSC+EC scenario. In this framework the optical--UV emission is
produced by synchrotron radiation from relativistic electrons.  While the
X-ray band lies on the rising low-energy side of the high-energy component and  is produced mostly by the SSC emission, with a contribution from the high-energy tail of the synchrotron emission at soft X-rays in the VHE and F1 flux states. The GeV emission, including the VHE emission in the VHE state, is dominated by EC scattering of external photons,

In contrast to the SSC-only and EC-only cases, the SSC+EC model successfully reproduces the broadband spectrum with physically more reasonable parameters. The derived electron Lorentz factors, magnetic field strengths, and Doppler factors lie within the range generally expected for powerful blazars, without requiring extreme particle energies. 

The adopted external photon temperatures, of order \(10^{3}\) K, suggest that the dominant seed-photon field is associated with the infrared torus rather than the BLR. This interpretation is also consistent with the detection of VHE photons, since strong \(\gamma\gamma\) absorption would be expected if the emission region were located inside the BLR.

Overall, the SSC+EC scenario provides the most self-consistent description of the broadband emission. It overcomes the main limitation of the SSC-only model, which requires extreme electron parameters, and the EC-only model, which fails to reproduce the VHE flux. Also the equipartition value was fixed at unity in all the flux states. The best-fitting model parameters and the corresponding reduced \(\chi^{2}\) values are listed in Table~\ref{tab:parameters_combined}, while the model SEDs for the different flux states are shown in Figure~\ref{fig:sed_states}. The derived jet powers are also given in Table~\ref{tab:parameters_combined}, with the largest values obtained for the brighter states, indicating a clear link between the source activity and jet energetics.
The fitted bulk Lorentz factor and jet power are generally higher in the brighter states. This suggests that the observed flux enhancements are driven primarily by changes in Doppler boosting, jet energetics and particle break energy.

\section{Summary}
\label{summary}

We have carried out a detailed multi-wavelength temporal and spectral study of the changing-look blazar B2~1420+32 using \emph{Fermi}-LAT, \emph{Swift}-XRT and \emph{Swift}-UVOT observations covering MJD~58818--60721. Our main results are as follows.

\begin{enumerate}

\item The source shows pronounced variability across all energy bands, with the maximum variability observed in the $\gamma$-ray regime. The one-day binned \(\gamma\)-ray light curve reveals a major flare around MJD~60488, during which the integrated photon flux reached
\((4.62 \pm 0.29)\times10^{-6}\) ph\,cm\(^{-2}\)\,s\(^{-1}\), corresponding to an enhancement of about 60 times relative to the average flux reported in the 4FGL-DR4 catalogue. During this flare, the photon index hardened to ($2.19 \pm 0.14$), indicating a shift of the high-energy emission toward larger photon energies.

\item Across the full light curve the flux--index anti-correlation is statistically significant but weak in amplitude, indicating that any global harder-when-brighter behaviour is not strongly pronounced in B2~1420+32.

\item The fractional variability shows a clear energy dependence. The \(\gamma\)-ray band displays the largest \(F_{\rm var}\), the optical and UV bands show substantial but smaller variability, and the X-ray band is comparatively less variable.

\item The \(\gamma\)-ray--optical/UV correlations are strong (\(\rho = 0.74\)--0.88), while the \(\gamma\)-ray--X-ray correlation is only
moderate (\(\rho = 0.44\)), indicating that the X-ray emission does not track the \(\gamma\)-ray variability as closely as the optical and UV emission.

\item The optical/UV and \(\gamma\)-ray maxima occur in different outbursts: the highest \emph{Swift}-UVOT \(V\)-band flux was recorded during the 2020 event and the highest \(\gamma\)-ray flux in 2024, showing that the relative amplitudes of the two spectral components are not preserved between outbursts.

\item Five activity states (VHE, F1, F2, F3 and F4) were identified from the Bayesian-block representation of the \(\gamma\)-ray light curve together with the availability of simultaneous multi-wavelength coverage. VHE, F1 and F4 are high-activity states, F3 is intermediate and F2 marks the lowest-flux interval. A log-parabola model is statistically preferred for the \(\gamma\)-ray spectra of the F1 and VHE states, while a power law is adequate for the remaining states.

\item The X-ray spectra are described by a log-parabola model in all states. Significant negative curvature is measured in  F1,VHE, F2 and F4 states; the F3 state curvature is consistent with zero within their uncertainties. The X-ray spectral slope at 1~keV varies between \(\alpha \sim 1.5\) and \(\sim 2.0\) across the states and does not follow a simple harder-when-brighter trend.

\item The broadband SEDs of all five states were modelled under three leptonic scenarios: pure SSC, pure EC, and a combined SSC+EC description. All three yield statistically acceptable fits, and the reduced \(\chi^2\) values alone do not discriminate between them.

\item The SSC-only scenario can formally reproduce the broadband spectra but requires physically disfavoured parameters: very large break Lorentz factors, a minimum Lorentz factor that varies by orders of magnitude between states and falls below unity in two of them, weak magnetic fields of \(B \sim 0.02\)--0.03~G, and large departures from equipartition, with the VHE state requiring \(\log(P_{\rm jet}/{\rm erg\,s^{-1}}) = 48.68\).

\item The EC-only scenario provides acceptable fits to the optical--GeV data and yields more reasonable break Lorentz factors, but returns sub-relativistic minimum Lorentz factors in all states, requires bulk Lorentz factors of \(\Gamma_b \sim 29\)--47, and fails to reproduce the VHE emission.

\item The SSC+EC model provides the most physically self-consistent description, with magnetic fields spanning \(B \sim 0.39\)--1.11~G, bulk Lorentz factors of \(\Gamma_b \sim 11\)--28.7, break parameters \(\xi_b \sim 0.04\)--0.09, and jet powers of \(\log(P_{\rm jet}/{\rm erg\,s^{-1}}) \sim 45.4\)--46.8.  In this framework
the optical--UV emission is produced by synchrotron radiation, while the
X-ray band lies on the rising low-energy side of the high-energy component
and is dominated by SSC emission. The GeV emission, including the VHE
emission in the VHE state, is dominated by EC scattering of external
photons.

\item The external seed-photon field has a characteristic temperature of \(\sim 10^{3}\) K, consistent with an infrared torus origin rather than the broad-line region.

\item The brighter states require larger bulk Lorentz factors and higher jet powers. The magnetic field is not simply anti-correlated with brightness: it is comparable across F1--F4 but drops sharply in the VHE state, indicating that changes in magnetization contribute to the observed variability in that state. The electron indices show no monotonic relation with flux.  The outbursts states (F1 and F4)
have nearly identical $\delta$, $B$ and $p$, and differ mainly in the break energy.

\end{enumerate}

\section{DISCUSSION}
\label{discussion}

\subsection{Energy-dependent variability}

The fractional-variability analysis across the light curves reveals a clear energy
dependence of the variability. The \(\gamma\)-ray band shows the largest
\(F_{\rm var}\), the optical/UV bands show substantial but smaller variability,
and the X-ray band is comparatively less variable. The strong
\(\gamma\)-ray--optical/UV correlations indicate a close connection between these
emission components, whereas the \(\gamma\)-ray--X-ray correlation is only
moderate, suggesting that the X-ray emission does not track the \(\gamma\)-ray
variability as closely. This behaviour is broadly consistent with the broadband
SED inferred from our modeling, in which the X-ray band mainly samples the
low-energy side of the high-energy component, while the optical/UV and
\(\gamma\)-ray bands are associated with more energetic electrons producing
synchrotron and inverse-Compton emission. The lower variability in the X-ray band
can therefore be understood in terms of lower-energy electrons with longer
cooling times, whereas the stronger optical/UV and \(\gamma\)-ray variability
likely reflects the more rapid evolution of higher-energy electrons in the jet.
Similar energy-dependent variability has been reported in other FSRQs, including
3C~279 and PKS~0903\(-\)57 \citep{2019MNRAS.484.3168S,2021MNRAS.504..416S}. A
comparable increase of \(F_{\rm var}\) with photon energy has also been reported
in HBLs \citep{2015A&A...576A.126A}; however, in those sources the X-ray band
usually traces the synchrotron peak itself, whereas in B2~1420+32 the X-ray band
appears to probe the onset of the high-energy component.The energy-dependent variability observed in B2~1420+32 therefore places important constraints on the underlying jet physics and is broadly consistent with leptonic scenarios in which the most energetic electrons dominate the strongest flux changes.

\subsection{X-ray spectral curvature}
The 0.3--10~keV spectra are better described by a log-parabola than by a
power law in four of the five activity states. The curvature parameter is
negative throughout these states, ranging from \(\beta = -0.16 \pm 0.01\) in
F4 to \(\beta = -0.38^{+0.06}_{-0.05}\) in F2, implying a concave continuum
that hardens toward higher energies.

Concave X-ray continua of this kind are well established in low- and
intermediate-synchrotron-peaked blazars, where the X-ray band lies near the
minimum between the two SED components. Broad-band \textit{BeppoSAX}
observations resolved the upturn directly in ON~231
\citep{2000A&A...354..431T}, S5~0716+714 \citep{1999A&A...351...59G} and
BL~Lacertae \citep{2002A&A...383..763R}, and a systematic
\textit{XMM-Newton} study of 14 LSP and ISP blazars recovered significant
negative log-parabolic curvature in half the sample, with \(\beta\) values
comparable to those measured here \citep{2018MNRAS.473.3638G}. In these
sources the concavity is attributed to the declining high-energy tail of the
synchrotron component giving way to the rising low-energy side of the
inverse-Compton component within the observed band. A similar picture has
emerged from broad-band studies of FSRQs in which the X-ray band contains
mixed synchrotron and Compton contributions
\citep{2019MNRAS.484.3168S,2021MNRAS.504..416S}.

We emphasize that a negative curvature parameter from a
log-parabolic fit does not, by itself, constitute proof that two
physically distinct emission components contribute significantly
within the XRT band; a single intrinsically curved component can
also produce $\beta < 0$. We therefore treat the curvature as
indicative of a complex X-ray spectral shape rather than as proof
of a spectral transition, and rely on the broadband SED modelling
(Section~\ref{broadband}) rather than the curvature parameter alone to assess the
relative synchrotron and inverse-Compton contributions.

\begin{figure*}
\centering

\begin{subfigure}[b]{0.4\textwidth}
\centering
\hspace{-0.8cm}
\includegraphics[width=0.9\linewidth,angle=-90]{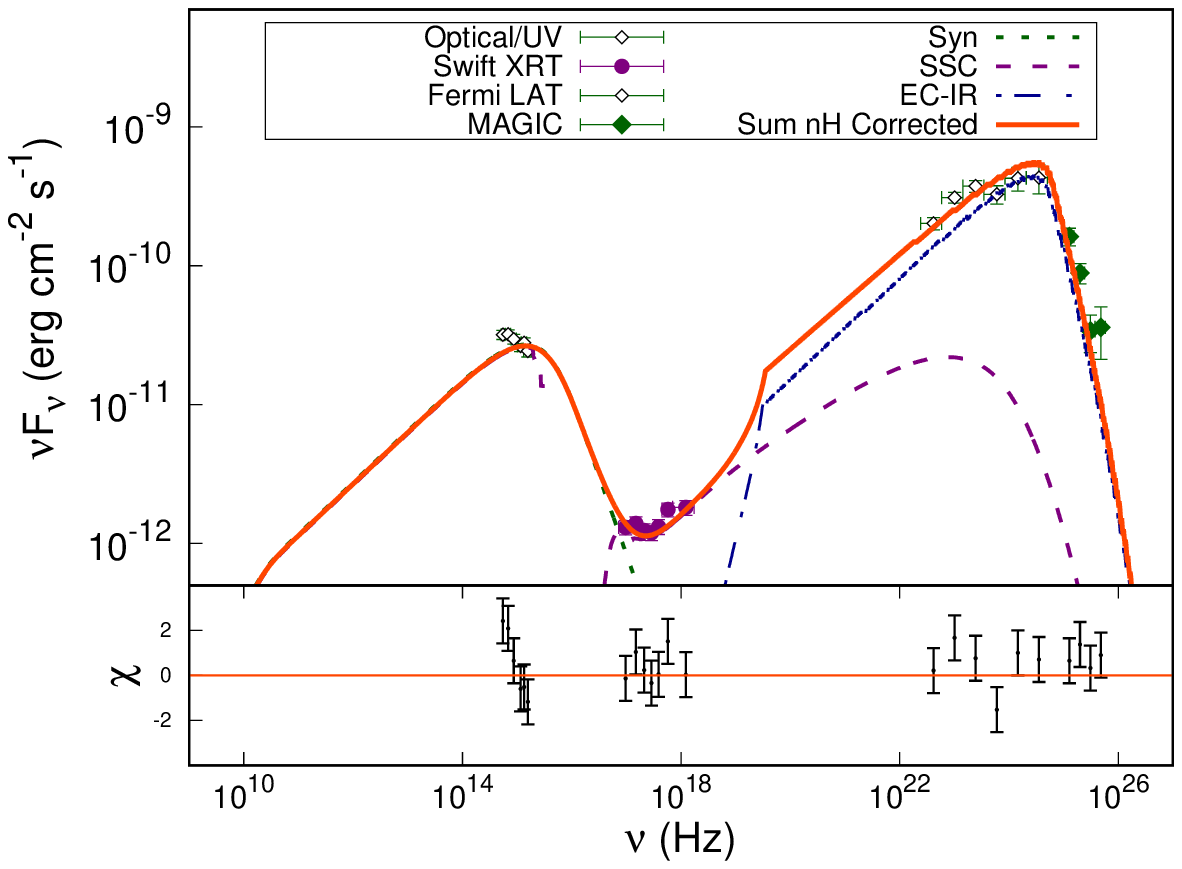}
\caption{VHE state}
\label{fig:sed_vhe}
\end{subfigure}
\hspace{1.1cm}
\begin{subfigure}[b]{0.4\textwidth}
\centering
\includegraphics[width=0.9\linewidth,angle=-90]{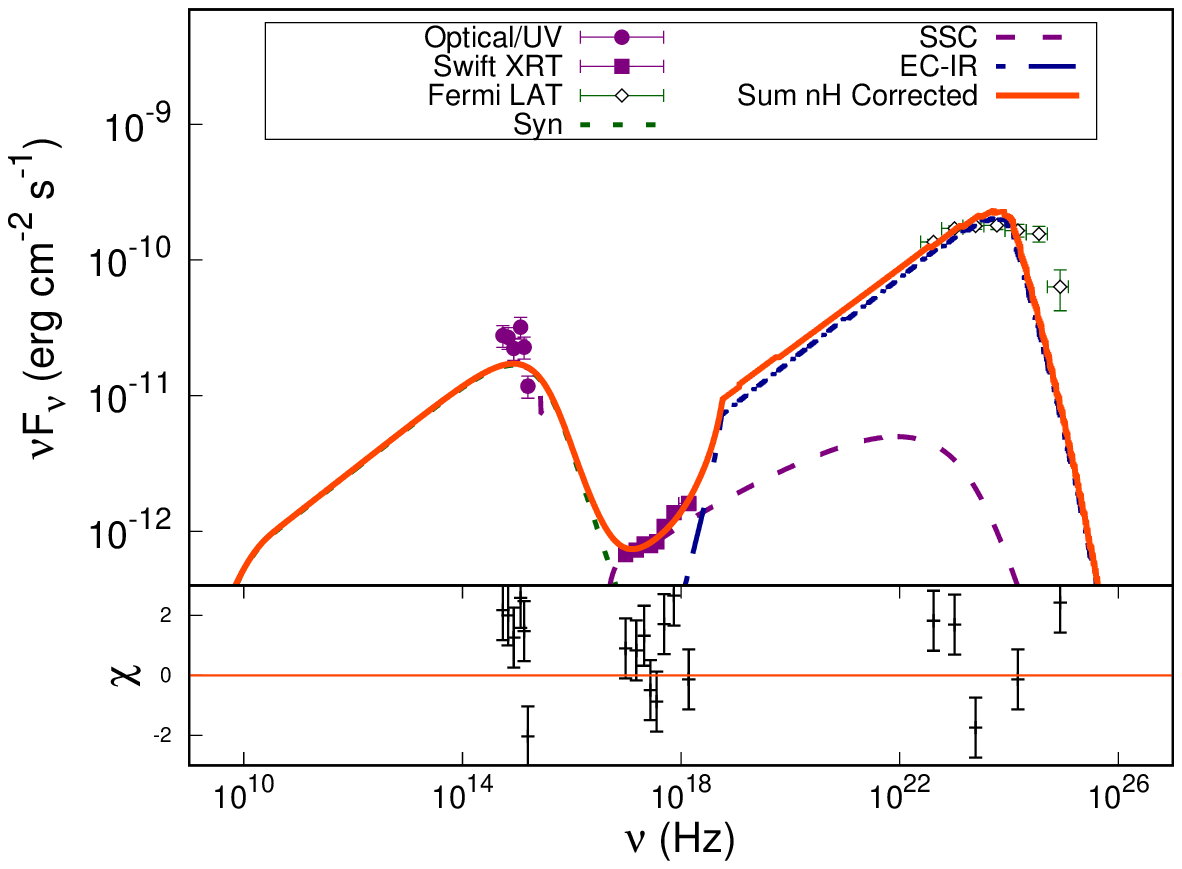}
\caption{F1 state}
\label{fig:sed_f1}
\end{subfigure}

\vspace{0.1cm}

\begin{subfigure}[b]{0.4\textwidth}
\centering
\hspace{-0.8cm}
\includegraphics[width=0.9\linewidth,angle=-90]{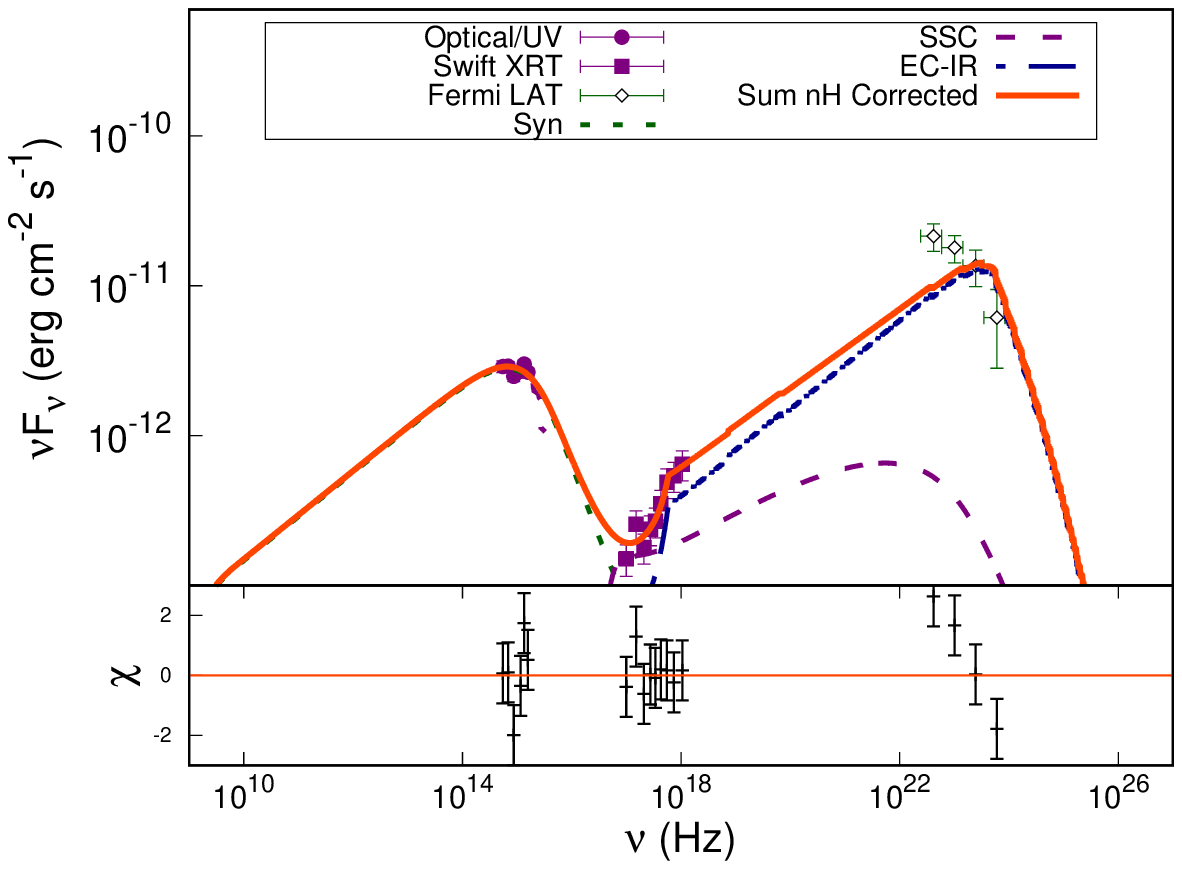}
\caption{F2 state}
\label{fig:sed_f2}
\end{subfigure}
\hspace{1.1cm}
\begin{subfigure}[b]{0.4\textwidth}
\centering
\includegraphics[width=0.9\linewidth,angle=-90]{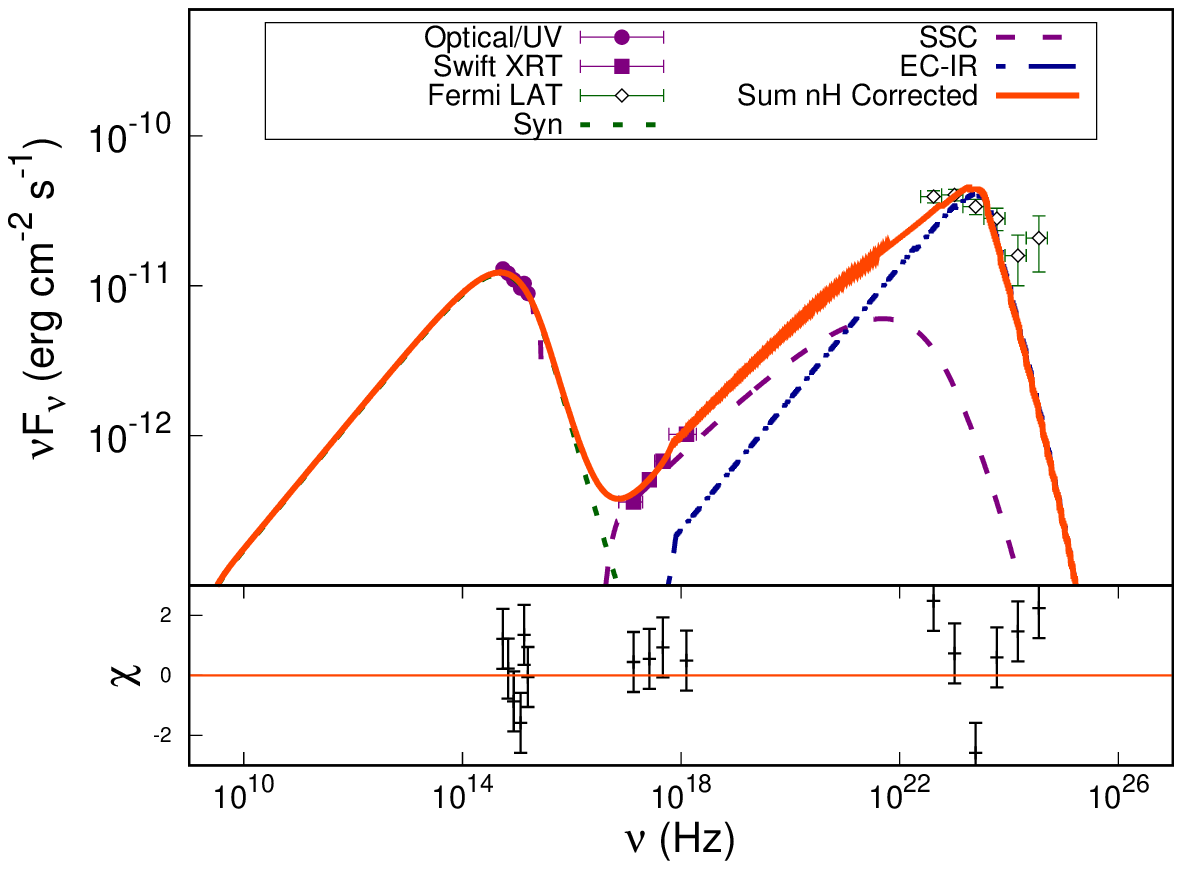}
\caption{F3 state}
\label{fig:sed_f3}
\end{subfigure}

\caption{Broadband spectral energy distributions (SEDs) of B2~1420+32 during different activity states. Panels show (a) VHE state, (b) F1 state, (c) F2 state, and (d) F3 state.}
\label{fig:sed_states}
\end{figure*}

\begin{figure*}
\centering

\begin{subfigure}[b]{0.35\textwidth}
\centering
\includegraphics[width=1\linewidth,angle=-90]{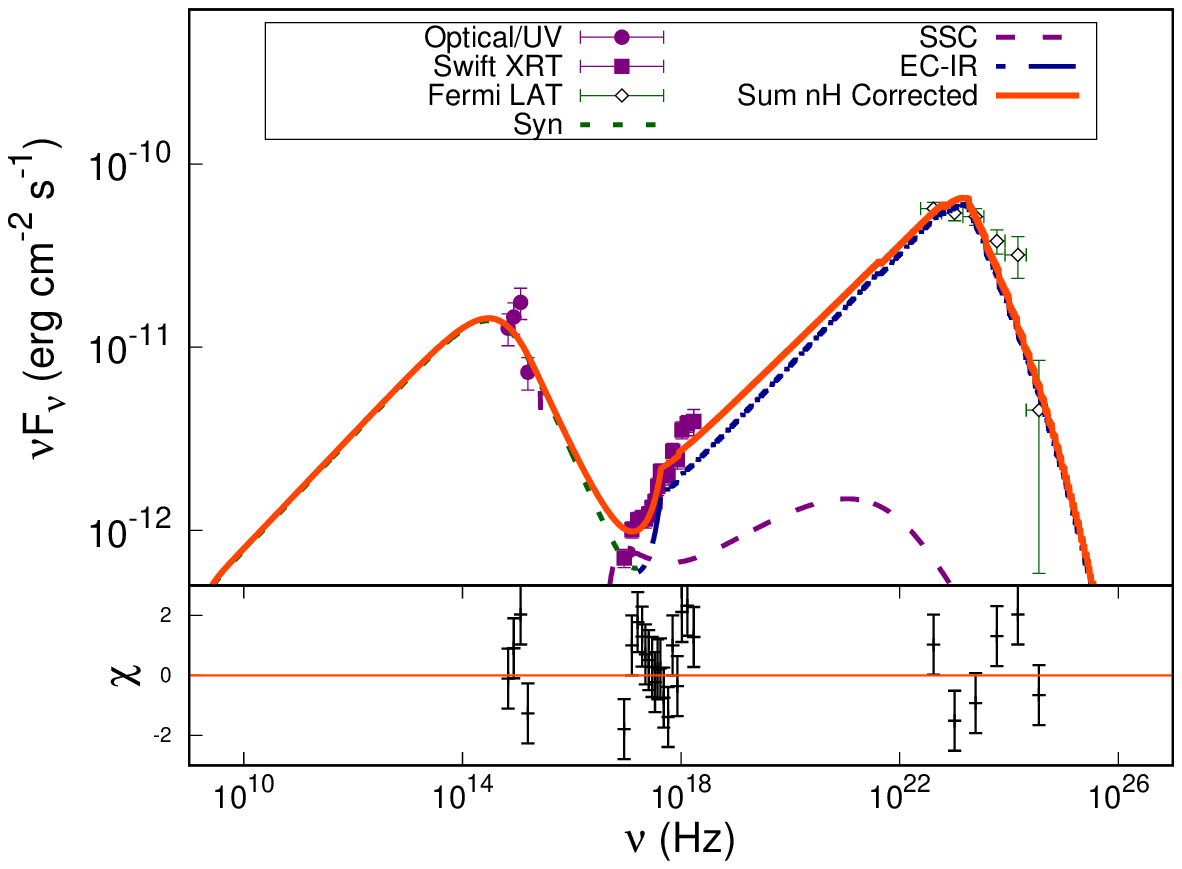}
\caption{F4 state}
\label{fig:sed_f4}
\end{subfigure}

\caption{Broadband spectral energy distribution (SED) of B2~1420+32 during the F4 state.}
\label{fig:sed_f4_full}
\end{figure*}

Our SED modelling supports the same interpretation. In all five states the
XRT points lie close to the minimum of the modelled SED, on the rising
low-energy side of the high-energy component (Figures~\ref{fig:sed_states} and \ref{fig:sed_f4_full}). In the VHE and F1 states, where
the model components are best separated, the high-energy tail of the
synchrotron component is still contributing at the softest energies. As
noted above, the concavity of the rising inverse-Compton component in this
band means the measured \(\beta\) need not indicate two components of
comparable strength; the SED modelling shows, however, that the X-ray band
consistently falls on the low-energy side of the high-energy hump rather
than on the synchrotron peak itself. This is consistent with the
comparatively low X-ray \(F_{\rm var}\) and the only moderate
\(\gamma\)-ray--X-ray correlation reported in Section~\ref{temporal}, both of which
indicate that the X-ray band does not share a single dominant origin with
the \(\gamma\)-ray emission. A related conclusion was drawn for the 2020
outburst of this source from the variation of the X-ray spectral index
across the flare \citep{2021A&A...647A.163M}.

The F3 state is the exception, returning \(\beta = 0.03^{+0.06}_{-0.05}\),
consistent with a simple power law. Since the curvature in this band depends
on the relative weighting of the two components and on the position of the
crossing point, the flat spectrum of F3 is naturally read as a state in which
one component dominated the 0.3--10~keV range more completely than in the
others. We caution that curvature at the level measured in F4 would be
difficult to detect in a spectrum of this quality, so the difference between
F3 and the remaining states should be regarded as suggestive rather than
established. Taken with the state-to-state variation of \(\alpha\), which is
softest in the VHE state (\(\alpha \simeq 1.97\)) and hardest in the low-flux
F2 state (\(\alpha \simeq 1.50\)) and so runs counter to a simple
harder-when-brighter trend, this indicates that the relative contributions of
the synchrotron and Compton components to the X-ray band change between
activity states, as expected in a source whose broad-band spectral shape is
not preserved from one outburst to the next. 

\subsection{State-dependent X-ray behaviour in the context of the broadband SED
modelling}

We identified five activity states (VHE, F1, F2, F3, and F4) based on the observed variability and the availability of simultaneous X-ray, optical/UV, and $\gamma$-ray observations. Across these states, the X-ray flux varies between \(\sim 1.8 \times 10^{-12}\) erg cm\(^{-2}\) s\(^{-1}\) and \(\sim 7.6 \times 10^{-12}\) erg cm\(^{-2}\) s\(^{-1}\). We noted that the VHE state, despite its enhanced \(\gamma\)-ray activity, does not coincide with the peak X-ray flux. This, together with the moderate $\gamma$-ray--X-ray correlation, indicates that the temporal evolution is not controlled by a single radiative component alone. Similar state-dependent multiwavelength variability patterns have been reported in other FSRQs and are commonly interpreted in terms of stratified or evolving physical conditions within the jet \citep{Boettcher:2006pd,2019MNRAS.484.3168S,2021MNRAS.504..416S}. Although the highest-flux state F4 shows a relatively hard spectrum, the VHE state is comparatively softer, and some intermediate or lower-flux states are nearly as hard as the brighter states. In the context of our SED modeling, this behaviour is consistent with the X-ray band sampling the transition between the high-energy synchrotron component and the low-energy inverse-Compton emission, with the relative contributions of these components varying from one flux state to another.

\subsection{Discriminating between the radiative scenarios}

All three scenarios considered here, pure SSC, pure EC, and a combined SSC+EC
description, provide statistically acceptable descriptions of the broadband data. Fit quality alone therefore cannot
select among them, and our preference for the SSC+EC description rests on the
physical acceptability of the inferred parameters.

Three features disfavour the SSC-only scenario, all set out in
Section~\ref{ssc_only}. The derived minimum Lorentz factor varies by orders of
magnitude between states and falls below unity in two of them, which is
unphysical and indicates that this parameter is absorbing model degeneracies
rather than tracking a genuine low-energy cutoff. The inferred break Lorentz
factors are more than an order of magnitude above the values typically obtained
in one-zone leptonic models of powerful FSRQ-like blazars, and would require
unusually efficient particle acceleration or comparatively weak radiative
cooling. The third difficulty is energetic: reproducing the GeV emission through
SSC alone requires a weak magnetic field and a correspondingly large electron
energy density, leaving the emitting region strongly particle dominated in every
state, and the VHE state requires a jet power some two orders of magnitude above
the accretion-disk luminosity inferred for this source from its broad-line
emission \citep{2021A&A...647A.163M}, and super-Eddington for any plausible black
hole mass. The remaining states are less extreme, so the energetic difficulty is
specific to the state in which VHE emission was detected.

The EC-only scenario performs better in some respects, as described in
Section~\ref{Ec_only}. It returns break Lorentz factors broadly consistent with
the values expected from radiative cooling in the presence of a strong external
photon field, and the departure from equipartition is smaller than in the
SSC-only case. The scenario nevertheless fails on three counts. It returns
sub-relativistic minimum Lorentz factors in all five states, corresponding to
electron energies below the rest-mass energy, so that the low-energy end of the
electron distribution is again poorly constrained. It requires bulk Lorentz
factors of \(\Gamma_{b} \sim 29\)--47, substantially larger than the value of
\(19 \pm 9\) measured for the parsec-scale jet of this source from VLBA
observations \citep{2021A&A...647A.163M}.  In addition, scattering of the
infrared external photons increasingly suffers from Klein--Nishina suppression
at VHE energies, reducing the scattering efficiency and makes it difficult for
the EC component alone to sustain the observed emission toward several hundred
GeV.

The combined description avoids these difficulties (Section~\ref{com}). The
optical/UV emission is produced by synchrotron radiation, the X-ray band lies on
the rising low-energy side of the inverse-Compton component, EC scattering on the
infrared field dominates the GeV emission and VHE emission in the VHE flux state. The resulting
parameters are moderate in every state: the minimum Lorentz factors remain
relativistic, the break Lorentz factors lie in the range expected for FSRQ-like
sources, and the jet powers of
\(\log(P_{\rm jet}/{\rm erg\,s^{-1}}) \sim 45.4\)--46.8 are typical of powerful
blazar jets. The bulk Lorentz factors of \(\Gamma_{b} \sim 11\)--28.7 are broadly
consistent with the VLBA measurement, with the VHE state at the upper edge of the
\(1\sigma\) range, in contrast to the substantially larger values required by the
EC-only fits. Additionally, in the SSC+EC scenario, the model parameters are obtained under the assumption of equipartition between the energy density of the relativistic particles and that of the magnetic field.

The inferred seed-photon temperature of \(\sim 10^{3}\) K favours an infrared
torus origin rather than the BLR. This is supported by the VHE detection itself,
since strong \(\gamma\gamma\) absorption would be expected if the emitting region
were located deep inside the BLR, and similar torus-seeded EC scenarios have been
invoked for other VHE-detected FSRQs
\citep{Costamante2018,vandenBerg2019}. The same conclusion was reached for the
2020 outburst of this source, where the emitting region was placed at a distance
well beyond the broad-line radius and within the dusty torus
\citep{2021A&A...647A.163M}. Our analysis extends that result to three further
activity states spanning 2021 to 2024, indicating that the torus-dominated
external field is a persistent rather than an episodic feature of this source.

\subsection{Jet energetics and the changing-look nature of the source}
 
The state-dependent parameter variations suggest that the dominant
changes are associated with jet energetics and Doppler boosting. The
bulk Lorentz factor varies from $\Gamma_b \sim 11$ in the lower-flux
states to $\sim 28.7$ in the VHE state, with the brighter VHE, F1 and F4
states generally requiring larger $\Gamma_b$ than F2 and F3. Relativistic
beaming therefore plays an important role in amplifying the observed
emission during active states, particularly in the VHE state. The
magnetic field is comparable across F1--F4 ($B \sim 0.93$--$1.11$\,G)
but drops sharply to $B \sim 0.39$\,G in the VHE state, indicating that
changes in magnetization contribute to the variability in that state
alongside changes in particle energization and bulk motion. The
low-energy electron index $p$ spans $\sim 2.1$--$2.5$ and the
high-energy index $q$ remains steep at $\sim 4.5$--$5.4$, but neither
shows a monotonic relation with flux: the hardest low-energy slope is
found in the intermediate F3 state rather than in the brightest state.
The break parameter $\xi_b$ shows significant variation and shows no
clear correlation with flux. The brighter states are therefore not
produced by a single universal change in the electron spectrum, but
reflect a combination of moderate changes in the particle distribution,
the magnetic field and the Doppler boosting.

The jet power shows the clearest systematic trend. The modelled values
span $\log(P_{\rm jet}/{\rm erg\,s^{-1}}) \sim 45.4$--$46.8$, with the
highest values in the bright VHE, F1 and F4 states and the lowest in F2
and F3. These are characteristic of powerful FSRQ jets and transition
blazars \citep{2008MNRAS.385..283C,10.1111/j.1365-2966.2011.18578.x},
and higher than those inferred for purely SSC-dominated BL~Lac states
\citep{2024MNRAS.527.5140S}. At the same time, both internal and
external seed-photon fields are required to describe the broadband
spectrum: SSC principally accounts for the X-ray emission, while EC
scattering on the infrared torus field accounts for the GeV emission and
extends to the VHE band. Neither process alone provides a physically
acceptable description (Sections 4.1 and 4.2). The source thus combines
FSRQ-like jet energetics with a radiative output in which an internal
photon field plays a significant role, consistent with B2~1420+32
occupying an intermediate regime between classical BL~Lac objects and
FSRQs \citep{Mishra_2021,10.1111/j.1365-2966.2011.18578.x}.

These results can be further understood in the context of the optical spectroscopic observations obtained during the 2020 outburst, the episode covered in full by our F1 state
(MJD\,58845--58911) and within which the VHE state (MJD\,58868--58872)
is nested. Three spectra were obtained with the LDT in response to the
MAGIC detection, at MJD\,58876.6, MJD\,58887.5 and MJD\,58904.4, that is
8, 19 and 36 days after the VHE flare of MJD\,58868.3--58870.3
\citep{2021A&A...647A.163M}. They give rest-frame Mg\,\textsc{ii}
equivalent widths of $2\pm1$, $6\pm3$ and $11\pm2$\,\AA\ respectively,
as the optical continuum faded by a factor of about five, while the
Mg\,\textsc{ii} line flux itself stayed constant within uncertainties.
The excursion across the conventional $EW = 5$\,\AA\ boundary was
therefore produced by the brightening of the jet continuum and not by a
change in the line-emitting gas. Our modelling identifies what drove
that continuum: the VHE state requires the largest bulk Lorentz factor
and the highest jet power of the five states, while the seed-photon
temperature remains at $\sim 10^3$\,K throughout. The external radiation
field is therefore essentially unchanged from state to state, and it is
Doppler boosting and jet energetics that make the continuum swing, which is precisely what is required to dilute the broad lines without altering them.

The comparison between these results should be interpreted in the context of the different time intervals over which the observations were obtained. Our F1
state is a single time-averaged SED built from all simultaneous
$\gamma$-ray, X-ray and optical/UV data between MJD\,58845 and
MJD\,58911, and the VHE state is a separate average over
MJD\,58868--58872.
The Mg\,\textsc{ii} equivalent widths, by contrast, come from three
individual spectra taken on MJD\,58876.6, MJD\,58887.5 and MJD\,58904.4,
during the decay of that outburst. No spectroscopic measurement exists within the VHE state window
itself, and the F1 SED averages over the whole interval across which the
equivalent width rose from $2$ to $11$\,\AA. We therefore compare
modelled parameters and measured equivalent widths that describe the
same outburst but not the same epochs within it, and the comparison
should be read as a consistency check between the two data sets rather
than as an epoch-by-epoch correspondence. A further limitation is that the present analysis does not measure emission-line properties and therefore cannot directly track the changing-look classification itself, which rests on
the spectroscopic monitoring of \citet{Mishra_2021} and
\citet{2021A&A...647A.163M}. What our modelling adds is a physical
account of the continuum behind that transition: its variations are
governed by Doppler boosting, jet power, particle break energy and magnetization rather than by changes in the external photon field.

\subsection{Comparison of the 2020 and 2024 outbursts}
\label{sec:2020vs2024}

Among the five activity states, F1 and F4 provide the most direct comparison
between two well-sampled episodes of enhanced activity, covering the 2020
outburst (MJD\,58845--58911) and the 2024 outburst (MJD\,60484--60530)
respectively. The two episodes do not order in the same way in the two bands.
The highest $V$-band flux of the entire campaign,
$(8.82\pm0.31)\times10^{-14}$ erg\,cm$^{-2}$\,s$^{-1}$\,\AA$^{-1}$, was
recorded at MJD\,58859 within F1, whereas the highest one-day $\gamma$-ray
flux, $(4.62\pm0.29)\times10^{-6}$ ph\,cm$^{-2}$\,s$^{-1}$, occurred at
MJD\,60488 within F4. Since both extrema are single epochs while the
corresponding SEDs are averaged over the full state intervals, the comparison
presented below refers to the mean broadband state of each outburst rather
than to its peak.

The model parameters obtained for the two states allow this difference to be
examined quantitatively. The peak frequencies of the individual components
are given by
\[
\nu_{\rm syn}=\frac{\delta}{1+z}\gamma_b^{2}\nu_B,\quad
\nu_{\rm SSC}=\frac{\delta}{1+z}\gamma_b^{4}\nu_B,\quad
\nu_{\rm EC}=\frac{\delta}{1+z}\gamma_b^{2}\Gamma_b\nu_{*},
\]
where $\nu_B=eB/2\pi m_ec$ is the Larmor frequency, $\nu_{*}=2.82k_BT/h$ is
the characteristic frequency of the external photon field, and
$\delta=[\Gamma_b(1-\beta\cos\theta)]^{-1}$ is evaluated at the fixed viewing
angle $\theta=2^{\circ}$. We obtain Doppler factors of $\delta=23.63$ and
$22.41$ for F1 and F4, and peak frequencies of
$\log(\nu_{\rm syn}/{\rm Hz})=15.12$, $\log(\nu_{\rm SSC}/{\rm Hz})=22.68$
and $\log(\nu_{\rm EC}/{\rm Hz})=23.65$ for F1, against $14.54$, $21.51$ and
$23.01$ for F4. All three components therefore peak at higher frequency
during the 2020 outburst. The Compton dominance, taking the summed SSC and EC
contributions as the high-energy component, is $11.17$ for F1 and $4.70$ for
F4.

The measured spectra show a related change. The X-ray spectra are described
by $\alpha=1.78\pm0.05$ and $\beta=-0.31\pm0.10$ in F1 and by
$\alpha=1.54^{+0.07}_{-0.01}$ and $\beta=-0.16\pm0.01$ in F4, so that the
2024 spectrum is both harder and less concave, as expected when the
synchrotron component peaks at lower frequency and contributes less within
the $0.3$--$10$\,keV band. The $\gamma$-ray indices, by contrast, are nearly
the same, $2.13\pm0.06$ in F1 and $2.20\pm0.05$ in F4, indicating that the
shift of the Compton peak is not large enough to alter the slope measured
across the LAT band.

These results also bear on the changing-look behaviour of the source. The
spectroscopic transition of the 2020 episode, during which the
Mg\,\textsc{ii} equivalent width fell below the conventional threshold,
occurred within the F1 interval, yet F1 carries the larger Compton dominance.
This is not a contradiction: the transition was captured at individual
epochs, whereas the F1 SED averages over the whole interval, so the brief
BL~Lac-like appearance is smoothed out. Averaged over the state, F1 lies
firmly in the Compton-dominated regime characteristic of FSRQs, more so than
F4. No comparable spectroscopy is available for the 2024 episode, so beyond
this point the two states can be compared only through their broadband
continua.

These results indicate that the physical conditions are not preserved even
among the high-flux states, and that comparable levels of activity may arise
from different combinations of parameters. Between F1 and F4 the Doppler
factor differs by 5 per cent, the magnetic field by 6 per cent and the bulk
Lorentz factor by 9 per cent, while the seed-photon temperature is unchanged
and the low-energy electron index is identical at $p=2.37$. The only
parameter to vary appreciably is the break Lorentz factor, which decreases
from $\gamma_b=5996$ to $3061$, a factor of $1.96$, and this alone accounts
for the measured shift of all three peaks. The dominant parameter responsible
for the difference between these two outbursts is therefore the break energy
of the electron distribution, with beaming and magnetization contributing
only marginally. It also accounts for the higher daily GeV photon flux of the 2024
event despite its lower Compton peak frequency, since the integral photon
flux above 100\,MeV is weighted towards the lowest energies of the band.

\subsection{Outlook}

Overall, the SSC+EC modeling suggests that the flux evolution of the source is governed mainly by changes in bulk motion, jet power, and the relative importance of internal and external seed-photon fields, rather than by a single parameter alone. B2~1420+32 has remained highly active since the interval analysed here, with renewed  flaring and a second detection at VHE energies at a flux comparable to that of the 2020 event. A recurrence of VHE emission at a similar level is a direct expectation of the picture developed in this work, in which the high-energy emission arises from a combination of SSC and external Compton scattering on an infrared torus field located beyond the broad-line region. B2~1420+32 therefore provides an important laboratory for investigating how
variations in jet energetics and seed-photon fields govern the broadband emission of changing-look blazars. Simultaneous VHE, X-ray and optical polarimetric coverage of a future outburst would provide a considerably stronger test of this picture than the present data allow, by constraining the magnetic-field geometry and the location of the emitting region independently of the SED fitting.

\section*{ACKNOWLEDGMENTS}

The authors thank the anonymous referee for valuable comments and suggestions. The authors extend their gratitude to the Inter-University Centre for Astronomy and Astrophysics (IUCAA), Pune, for providing support and facilities during the course of this work. We also express our gratitude to the Fermi-LAT collaboration for providing comprehensive long-term $\gamma$-ray data of high-energy sources. We acknowledge the HEASARC software packages, maintained by the NASA Goddard Space Flight Center, which were used in the reduction and analysis of the Swift-XRT and Swift-UVOT data.

\begin{table*}
\centering
\small

\caption{Best-fit model parameters for B2~1420+32 obtained by fitting the broadband SED model to the VHE, F1, F2, F3, and F4 states considering SSC+EC process. The table is organized into three sections. The top section lists the free parameters varied during the fit: the bulk Lorentz factor of the electron energy distribution ($\Gamma_b$), the magnetic field strength ($B$) in units of G, the broken power-law indices $p$ and $q$, and $\xi_b$. The upper middle section includes fixed parameters: $\xi_{\min}$ (in units of $10^{-4}$) and $\xi_{\max}$ and target photon temperature in K. The lower middle section includes the converted parameters $\gamma_{\min}$, $\gamma_b$, and ($\gamma_{\max}$ in units of $10^{6}$), derived from $\xi$. The bottom section lists $\log P_{\rm jet}$ (in units of erg,s$^{-1}$) and $\chi^2/\mathrm{dof}$. The emission region radius was fixed at $R = 10^{16}$,cm, and the jet inclination angle was fixed at $\theta = 2^\circ$. The parameters $\xi_b$, $\xi_{\min}$, and $\xi_{\max}$ represent the break, minimum, and maximum electron energies, respectively, with all energies expressed in units of $\sqrt{\mathrm{keV}}$. The equipartition parameter was fixed at unity ($B_{\rm eq}=1$) for all flux states.}

\label{tab:parameters_combined}

\setlength{\tabcolsep}{8pt}
\renewcommand{\arraystretch}{1.35}

\begin{tabular}{lccccc}
\hline
\textbf{Parameter} & \textbf{VHE} & \textbf{F1} & \textbf{F2} & \textbf{F3} & \textbf{F4} \\
\hline

\multicolumn{6}{l}{\textit{Free parameters}} \\
$\Gamma_{b}$ & $28.74_{-2.92}^{+4.11}$ & $15.10_{-0.93}^{+0.97}$ & $10.99_{-2.60}^{+10.20}$ & $11.65_{--}^{--}$ & $13.82_{-1.00}^{+2.24}$ \\
$B$ & $0.393_{-0.013}^{+0.011}$ & $0.932_{-0.015}^{+0.013}$ & $1.109_{-0.227}^{+0.538}$ & $1.042_{-0.059}^{+0.085}$ & $0.990_{-0.208}^{+0.147}$ \\
$p$ & $2.25_{-0.01}^{+0.01}$ & $2.37_{-0.03}^{+0.02}$ & $2.51_{-0.05}^{+0.05}$ & $2.11_{-0.01}^{+0.01}$ & $2.37_{-0.01}^{+0.02}$ \\
$q$ & $5.07_{-0.20}^{+0.34}$ & $5.01_{-0.40}^{+0.27}$ & $5.23_{--}^{--}$ & $5.41_{-0.38}^{+0.82}$ & $4.52_{-0.25}^{+0.36}$ \\
$\xi_{b}$ & $0.092_{-0.006}^{+0.001}$ & $0.080_{-0.010}^{+0.010}$ & $0.067_{-0.013}^{+0.032}$ & $0.054_{-0.006}^{+0.004}$  & $0.041_{-0.003}^{+0.012}$ \\

\hline
\multicolumn{6}{l}{\textit{Fixed parameters}} \\
$\xi_{\min}$ & 2.0 & 1.6 & 1.2 & 1.1 & 1.7\\
$\xi_{\max}$ & 15 & 253 & 13.5 & 3.18 & 15 \\
$T$ & 1096 & 1000 & 1000  &  860 &   1000   \\
\hline

\multicolumn{6}{l}{\textit{Converted parameters}} \\
$\gamma_{\min}$ & 20.95 & 11.9 & 9.2 & 8.5 & 12.7 \\
$\gamma_{\rm b}$ & 9641.51 & 5995.64 & 5114.77 & 4162.80 & 3061.20 \\
$\gamma_{\max}$ & 1.57 & 18.96 & 1.03 & 0.245 & 1.12 \\
\hline

$\log P_{\rm jet}$ & 46.79 & 46.54 & 45.44 & 45.62 & 46.72  \\
$\chi^{2}/{\rm dof}$ & 77.86/46 & 253.62/137 & 24.32/15 & 93.25/80 & 98.23/89 \\
\hline
\end{tabular}

\end{table*}

\bibliographystyle{aasjournal}
\bibliography{main_bib}






\bibliography{main_bib}{}
\bibliographystyle{aasjournalv7}

\begin{table*}[!t]
\centering
\footnotesize
\setlength{\tabcolsep}{12pt}          
\renewcommand{\arraystretch}{1.60}    

\caption{Details of the \textit{Swift}-XRT/\textit{UVOT} observations of B2~1420+32. The table is presented in two side-by-side sections for compactness. Columns (1) and (5): Observation ID; Columns (2) and (6): Observation date; Columns (3) and (7): XRT exposure time (s); Columns (4) and (8): UVOT exposure time (s).}
\label{tab:b2_obs}

\begin{tabular}{cccc@{\hspace{18pt}}cccc}
\hline
Obs. ID & Date & XRT (s) & UVOT (s) &
Obs. ID & Date & XRT (s) & UVOT (s)\\
\hline

00010520014 & 2020-01-02 & 2504.86 & 2461.78 &
00010520044 & 2023-04-18 & 1475.81 & 1426.24 \\

00010520015 & 2020-01-05 & 1673.18 & 1647.28 &
00010520045 & 2023-04-19 & 1469.34 & 1441.79 \\

00010520016 & 2020-01-08 & 1887.26 & 1839.11 &
00010520047 & 2023-04-20 & 1030.63 & 1006.60 \\

00010520017 & 2020-01-11 & 1990.79 & 1950.70 &
00010520048 & 2023-04-21 & 987.87 & 963.09 \\

00010520018 & 2020-01-19 & 1632.26 & 1626.97 &
00010520049 & 2023-04-22 & 834.93 & 810.66 \\

00010520019 & 2020-01-21 & 1551.45 & 1696.39 &
00010520050 & 2023-04-23 & 753.20 & 727.25 \\

00010520022 & 2020-01-24 & 1791.44 & 1697.76 &
00010520051 & 2023-04-24 & 950.26 & 922.54 \\

00010520023 & 2020-01-25 & 313.32 & -- &
00010520052 & 2023-04-24 & 561.63 & 535.11 \\

00010520024 & 2020-01-25 & 1820.29 & 1721.30 &
00010520053 & 2023-04-25 & 804.86 & 781.12 \\

00010520025 & 2020-01-27 & 1812.87 & 1896.36 &
00010520054 & 2023-04-26 & 737.14 & 712.98 \\

00010520026 & 2020-01-28 & 2041.79 & 1966.39 &
00010520055 & 2023-04-27 & 882.57 & 856.47 \\

00010520027 & 2020-01-31 & 1236.13 & 1223.30 &
00010520056 & 2023-05-10 & 955.44 & 929.08 \\

00010520028 & 2020-02-01 & 1687.42 & 1667.98 &
00010520057 & 2023-05-12 & 802.34 & 777.89 \\

00010520029 & 2020-02-05 & 2033.47 & 1983.58 &
00010520058 & 2023-05-14 & 1075.63 & 1050.25 \\
00010520030 & 2020-02-10 & 2427.05 & 2422.97 &
00010520059 & 2023-05-16 & 892.60 & 866.13 \\

00010520031 & 2020-02-13 & 178.01 & 177.78 &
00010520060 & 2023-05-18 & 932.83 & 908.77 \\

00010520032 & 2020-02-18 & 1813.53 & 1812.95 &
00010520061 & 2023-05-20 & 905.14 & 878.33 \\

00010520033 & 2020-02-19 & 2146.29 & 2145.10 &
00010520062 & 2023-05-22 & 932.71 & 908.23 \\

00010520034 & 2020-02-22 & 2349.33 & 2345.63 &
00010520063 & 2023-05-24 & 774.75 & 750.44 \\

00010520035 & 2021-08-12 & 2133.89 & 2041.69 &
00010520064 & 2024-06-28 & 1281.21 & 1256.23 \\

00010520036 & 2021-08-19 & 1842.95 & 1792.21 &
00010520065 & 2024-06-30 & 1373.99 & 1323.41 \\

00010520037 & 2021-08-26 & 1870.44 & 1819.56 &
00010520066 & 2024-07-07 & 1476.80 & 1451.21 \\

00010520039 & 2023-04-13 & 2053.48 & 2002.14 &
00010520067 & 2024-07-10 & 1226.06 & 1174.78 \\

00010520040 & 2023-04-14 & 1677.39 & 1649.50 &
00010520068 & 2024-07-13 & 1293.77 & -- \\

00010520041 & 2023-04-15 & 1727.53 & 1697.16 &
00010520069 & 2024-07-18 & 1449.22 & 1421.44 \\

00010520042 & 2023-04-16 & 1298.78 & 1271.04 &
00010520070 & 2024-07-19 & 1148.44 & 1115.47 \\

00010520043 & 2023-04-17 & 1634.76 & 1609.89 &
00010520072 & 2024-07-31 & 1602.16 & 1546.68 \\

\hline
\end{tabular}
\end{table*}



\end{document}